\documentclass [pre]{revtex4}
\usepackage{graphicx}
\usepackage{epsfig}

\newcommand{\Keywords}[1]{\par\noindent 
{\small{\em Keywords\/}: #1}}

\begin{document}

\title{{\itshape {\bf The Ghost of Stochastic Resonance: An Introductory Review}}}
 \thanks{To appear in Contemporary Physics (2011).}

\author{Pablo Balenzuela$^{a,b}$, Holger Braun$^{c,d}$ and Dante R. Chialvo$^{a,e,f}$   
\\
$^{a}${\em{Consejo Nacional de Investigaciones Cient\'ificas y Tecnol\'ogicas (CONICET), Buenos Aires, Argentina}};
$^{b}${\em{Departamento de F\'isica, Facultad de Ciencias Exactas y Naturales, Universidad de Buenos Aires, Buenos Aires, Argentina}};
$^{c}${\em{Heidelberg Academy of Sciences and Humanities, Heidelberg,Germany}};
$^{d}${\em{Centre for Ice and Climate, Niels Bohr Institute, University of Copenhagen, Copenhagen, Denmark}};
$^{e}${\em{Department of Physiology, David Geffen School of Medicine, UCLA, Los Angeles, CA, USA}};
$^{f}${\em{Facultad de Ciencias M\'edicas, Universidad Nacional de Rosario, Rosario, Santa F\'e, Argentina.}}\vspace{20pt}}

\begin{abstract}
Nonlinear systems driven by noise and  periodic forces with more than one frequency exhibit the phenomenon of {\it Ghost Stochastic Resonance} (GSR) found in a wide and disparate variety of fields ranging from biology to geophysics. The common novel feature is the emergence of a ``ghost" frequency in the system's output which it is absent in the input. As reviewed here, the uncovering of this phenomenon helped to understand a range of problems, from the perception of pitch in complex sounds or visual stimuli, to the explanation of climate cycles. Recent theoretical efforts show that a simple mechanism  with two ingredients are at work in all these observations. The first one is the linear interference between the periodic inputs and the second a nonlinear detection of the largest constructive interferences, involving a noisy threshold. These notes are dedicated to review the  main aspects of this phenomenon, as well as its different manifestations described on a bewildering variety of systems ranging from neurons, semiconductor lasers, electronic circuits to models of glacial climate cycles.\vspace{20pt}
\end{abstract}
\maketitle

\section{Introduction}

\subsection{Beyond Stochastic Resonance.}

The phenomenon of Stochastic Resonance (SR) \cite{Benzi1,Benzi2, Longtin, SRMoss,SR1,SR2,SR3}, in which an optimal level of noise allows a nonlinear system to follow the periodicity of the input signal, received ample attention in a number of disciplines over the last decades. The term Stochastic Resonance was introduced by Benzi, Parisi, Sutera and Vulpiani \cite{Benzi1,Benzi2} in their original conjecture explaining the earth's ice ages periodicities. The word resonance referred to the existence of a maximum in the system's response synchronous with the periodic input, observed for a certain amplitude of the noise. 

Initial theoretical work to understand SR on discrete or continuous bistable models emphasized the non-linear  aspects of the phenomenology \cite{SR1,SR2,SR3}, while other efforts \cite{dykman1, dykman2} noted that its main aspects can be described within the framework of linear response theory, such as for example the relation between the sinusoidal input and the spectral output component of the same frequency.
While SR was initially considered to be restricted to the case of periodic input signals, now it is widely used  including aperiodic or broadband input signals. Differences exist also in current measures of the system's output performance: the signal-to-noise ratio (SNR) is used for periodic inputs, mutual information for  random or aperiodic signals and linear correlation between output and input signals in other cases.

This introductory review is dedicated to a variant of SR termed Ghost Stochastic Resonance (GSR) which is ubiquitous for nonlinear systems driven by noise and periodic signals with {\it more than one frequency}.  GSR was first proposed \cite{dante1, dante2}  to explain how a single neuron suffices to detect the periodicities of complex sounds.  Later, similar dynamics was found to describe the quasi-periodicity observed in  abrupt temperature shifts during the last ice age,  known as Dansgaard-Oeschger (DO) events \cite{braun1}. Between these two examples, other manifestations of GSR were subsequently identified in disparate systems including neurons, semiconductors lasers, electronic circuits, visual stimuli, etc.  In the following sections GSR will be reviewed including its main underlying ingredients and the most relevant manifestations in different natural phenomena.

\subsection{Two examples of unexplained periodicities.}

To introduce the main aspects of GSR, we will use two examples where a system responds with spectral components which are not present in their inputs.  

\subsubsection{The missing fundamental illusion in pitch perception}\label{IntroPitch}

Pitch is a subjective attribute by which any sound can be ordered on a linear scale from low to high. If a tone is composed of a single frequency, the perceived pitch is its frequency. In  the case of complex sounds, composed of several pure tones, there is no an objective measure of the pitch, despite the fact that the accuracy of different observers to distinguish between different complex sounds can be as small as a few percents, even for untrained ears.   

\begin{figure}[htpb]
\epsfysize=10cm
\begin{center}
\epsfbox{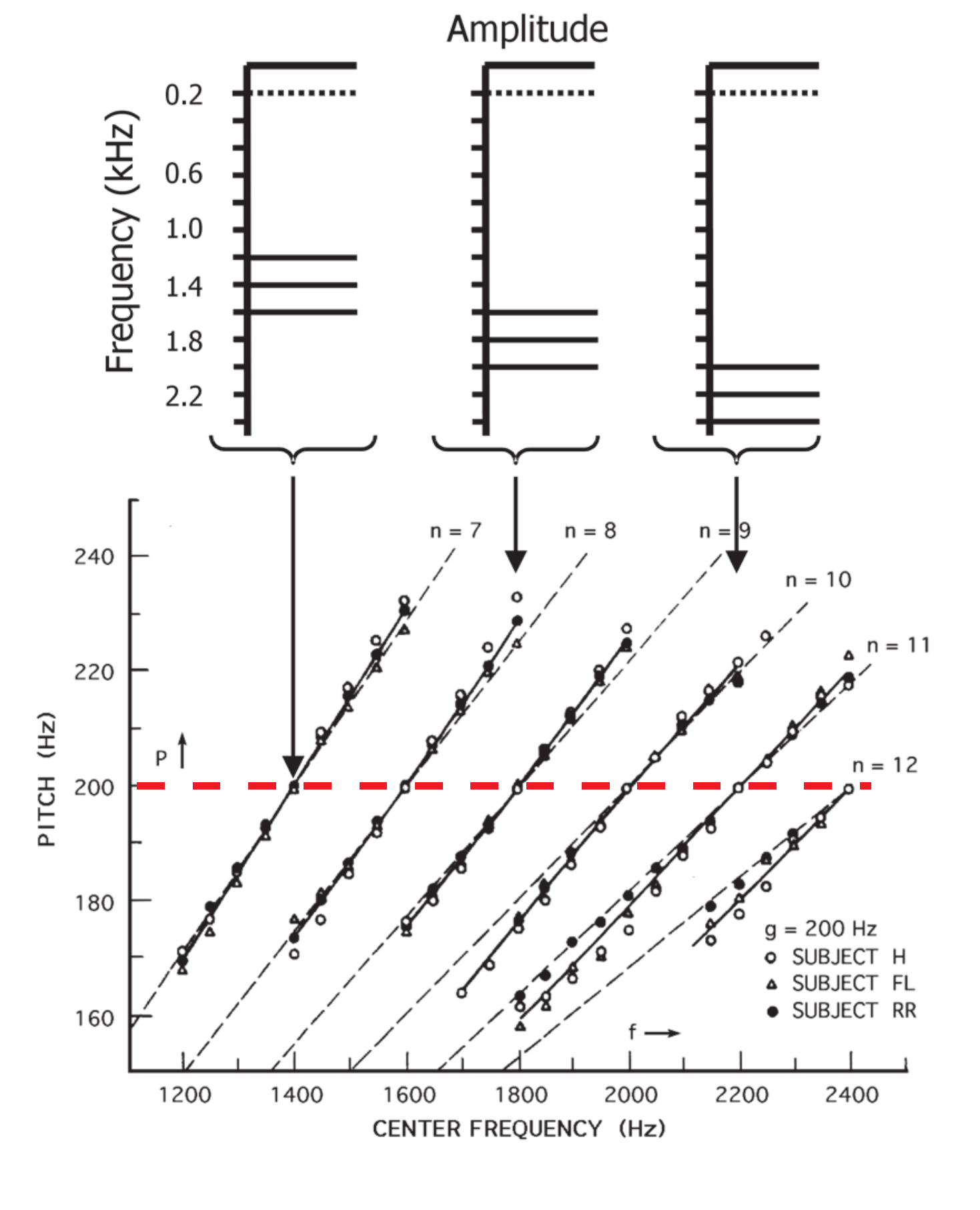}
\caption{\label{fig_schou} Results from Schouten's experiments  demonstrating that equidistant tones do not produce a constant pitch. The authors used a complex sound as described in Eq.\ref{SAM} with center values  from $1.2$ to $2.4$~kHz. Top diagrams depict three examples of the  frequency spectra of the complex sound used (with center frequencies of $1.4$, $1.8$ and $2.2$~kHz). Dotted lines correspond to the missing component $g$. The bottom graph indicates, with symbols (open or filled circles and triangles), the pitch heard by the three subjects for each complex sound. The dashed lines show that a $1/n$ function consistently underestimates the linear relation between the frequency and pitch shift. 
Reprinted with permission from  D.R. Chialvo, Chaos 13, pp. 1226-30 (2003). Copyright (2003) by the American Institute of Physics.
} 
\end{center}
\end{figure}

A well known illusion used to study pitch perception takes place when two tones of different frequencies are heard together. The paradox is that, under these conditions, the perception corresponds to a third, lower pitched, tone and not to any of the two frequencies. This is referred to as the {\it missing fundamental illusion} because the perceived pitch corresponds to a fundamental frequency  for which there is no actual source of air vibration. 
A characteristic phenomenon in  pitch perception is the so-called {\it pitch shift}. This refers to the variation of the perceived pitch when the frequencies of an harmonic tone are rigidly displaced. The first quantitative measurements of this phenomena are  reproduced in Fig.\ref{fig_schou} (\cite{DeBoer} and \cite{schouten}). 

The experiment is described as follows:  A sinusoidal amplitude modulated sound:

\begin{equation}
s(t)=(1+\cos(2\pi g t))\sin(2\pi f t) = \frac{1}{2}\sin(2\pi(f-g)t)+\sin(2\pi f t)+ \frac{1}{2}\sin(2\pi(f+g)t),
\label{SAM}
\end{equation}

is presented to a group of subjects which were asked to report the perceived pitch. This is a complex tone composed by three components equi-spaced in frequency by a value $g$. If $f=ng$ with $n$ integer, the three tones are higher harmonics of the fundamental $g$. If  the three components are displaced by $\Delta f$, e.g, $(f=ng+\Delta f)$,  the three tones are no longer higher harmonics of $g$ ($g$ is not the missing fundamental) but the difference between them remains equal to $g$.  These stimuli are represented in the top diagram of  Fig.~\ref{fig_schou} for $g=200$~Hz,  $f=1.4$~kHz and $\Delta f$ between $0$ and $1$~kHz. 

It was assumed for many years that, under these experimental conditions, the auditory system would report a pitch corresponding to nonlinear distortions \cite{helmholtz}. In other words, pitch perception would be related to the difference between the intervening tones. If that were the case, the reported pitch should remain constant (i.e., the red dashed line of bottom panel of Fig.~\ref{fig_schou}). Instead, the  reported pitches fall along straight lines of slopes close (but not exactly) to $1/n$.  In addition, there is a notorious ambiguity in the judgement, which is peculiar in the sense that although the same stimulus  may produce different  percepts they group in well defined values of pitch. The results in section \ref{SecPitch} will show that all of the quantitative and qualitative aspects of these experiments can be replicated at once in terms of Ghost Stochastic Resonance. 

\subsubsection{The Dansgaard-Oeschger events}\label{IntroGlacial}
\begin{figure}[htpb]
 
\begin{center}
\includegraphics[width=4in]{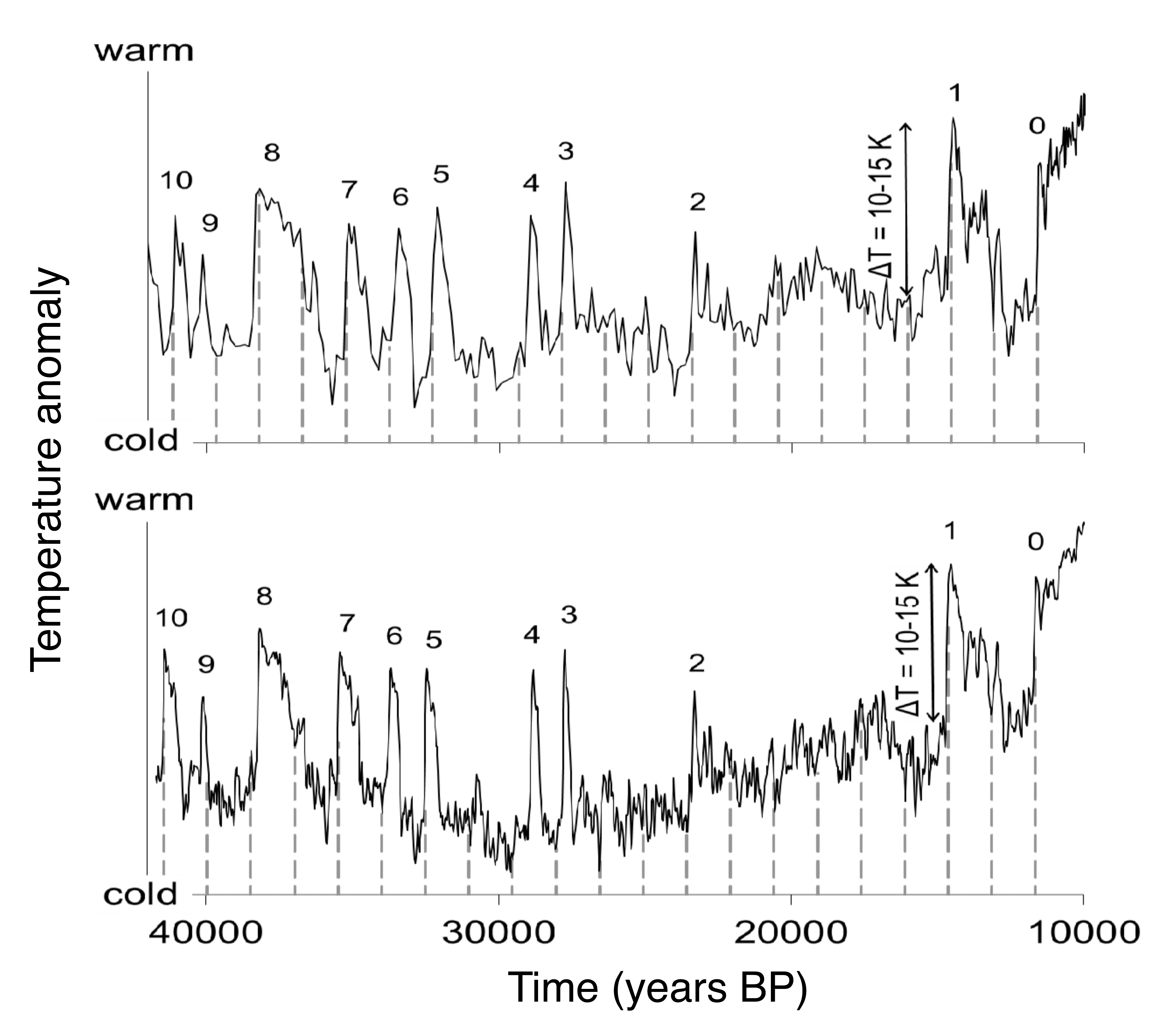}
\end{center}
\caption{\label{FigO18} Greenland's temperature anomalies during the time interval between $42.000$ and $10.000$ years before present (BP) as reconstructed from the ratio of two stable isotopes as measured in two deep ice cores from Greenland, the GISP2 (top) and NGRIP (bottom)
ice cores. Numbers label the Dansgaard-Oeschger events 0-10, following standard
paleoclimatic convention. Dashed lines are spaced by approximately 1.470 years. Note that
many events recur almost exactly in near-multiples of 1.470 years, e.g. the events
0, 1, 2, 3, 5, 7 and 10 as recorded in the GISP2 data. Both ice core records are shown on their
respective standard time scale, which was constructed by independent counting of annual
layers in the ice cores. Data from \cite{braun2,braun5a,gisp2}.
} 
 
\end{figure}

 Many paleoclimatic records from the North Atlantic region and Eurasia show a pattern of rapid climate oscillations, the so-called Dansgaard-Oeschger (DO) events \cite{braun1,braun2,braun3,braun4,braun5,braun5a}, which seem to exhibit a characteristic recurrence time scale of about $1470$~years during the second half of the last glacial period \cite{braun6,braun6a,braun6b}. Fig.~\ref{FigO18} shows the temperature anomalies during the events as reconstructed from the ratio of two stable oxygen isotopes as measured in two deep ice cores from Greenland, the GISP2 (Greenland Ice Sheet Project 2 \cite{braun2,gisp2}) and NGRIP (North  Greenland Ice Core Project \cite{braun5a}) ice cores, during the interval between $10000$ and $42000$ years before present (BP). Note that this isotopic ratio is a standard indicator to infer about temperature variations in many paleoclimatic records. The numbers $0-10$ in Fig.~\ref{FigO18} label the Dansgaard-Oeschger events, many of which are almost exactly spaced by intervals of about $1470$~years or integer multiples of that value.     


It has often been hypothesized that solar variability could have played a role
in triggering these rapid temperature shifts. However, whereas many solar and
solar-terrestrial records, including the historical sunspot record, were
reported to exhibit cycles of about $88$ and $210$ years, no noteworthy
spectral component of about $1470$~years has been identified in these records
\cite{braun7a,braun7b,braun8,braun9,braun10,braun11,braun12}. A turning
point in this discussion is the work of Braun et al. \cite{glacial}, showing
that an ocean-atmosphere model can even generate perfectly periodic
Dansgaard-Oeschger-like output events, spaced exactly by $1470$~years or
integer multiples of that value, when driven by input cycles of about 87 and 210
years. In other words, no input power at a spectral component corresponding to
$1470$~years is needed to generate output events with maximum spectral power at
that value. This work indicates that the apparent $1470$~year response time of
the events could result from a superposition of two shorter input cycles,
together with a strong nonlinearity and a millennial relaxation time in the
dynamics of the system as another manifestation of the GSR, as will be
discussed in section \ref{sec:D-O}. 

It is important to notice that the {\it stochastic aspect of the GSR} is relevant to replicate the entirety of the experimental observations of pitch perception and DO events described above. However, from a theoretical point of view it can be possible to have similar scenarios with a deterministic or suprathreshold ghost resonance, or also to replace the noise by a high frequency signal, as will be discussed in the last section.

\section{Ghost Stochastic Resonance}

\begin{figure}
\begin{center}
\includegraphics[width=4in]{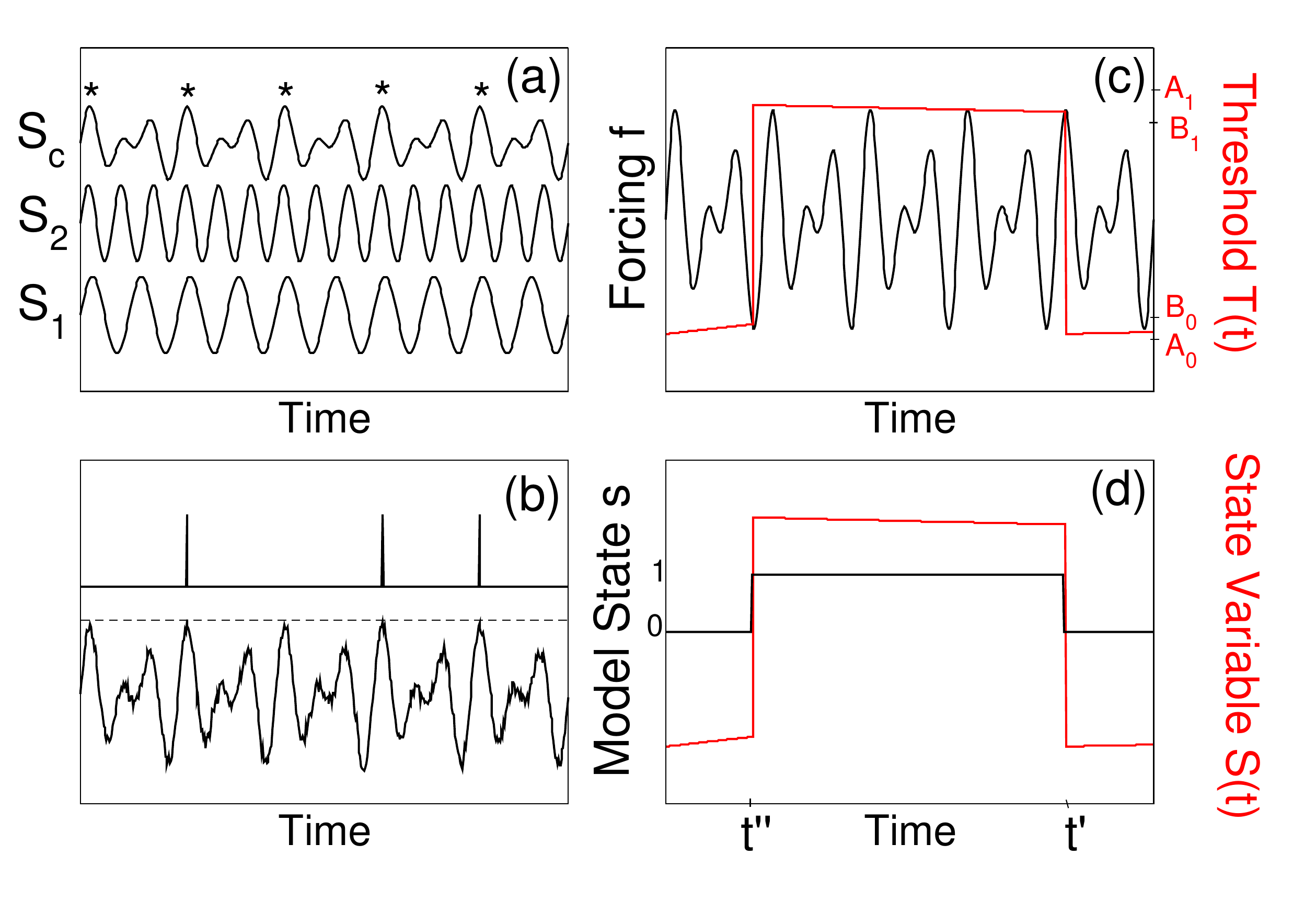}
\end{center}
\caption{Two examples of  GSR toy models. Left panels: Robust nonlinear stochastic detection of the missing fundamental $f_0$ by a constant  threshold. (a) Example of a complex sound $S_c$ built by adding two
  sinusoidal $S_1$ and $S_2$ frequencies: $f_1=kf_0$ (bottom) and
  $f_2=(k+1)f_0$ (middle). Specifically, $x(t)=\frac{1}{2}(A_1\sin(f_12\pi
  t)+A_2\sin(f_22\pi t))$  with $A_1=A_2=1$ ; $k=2$ and $f_0=1$Hz in this
  case. The peaks (asterisks) exhibited by $S_c$ result from the constructive
  interference of $S_1$ and $S_2$ are spaced at the period of the missing fundamental
  $f_0$. (b) The peaks of the signal shown in (a) can be reliably detected by
  a noisy threshold, generating inter-spike intervals close to the fundamental
  period or to an integer multiple thereof. Right panels: Conceptual climate model with a non-constant bi-threshold function as described in Braun et al \cite{braunchialvo} . Panel(c):  Complex signal $S_c$ as shown in panel (a) and the bi-threshold exponential function (in red). Panel (d): the same bi-threshold exponential function and  the state function (with value 1 if the system is in warm state and 0 if it is in the cold state) for the conceptual climate model. }
\label{Fig1}

\end{figure}

The two examples described above can be modeled in terms of GSR. Despite some  flavors  the models are build upon two basic ingredients:  linear interference of pure tones plus a threshold,  which plays the role of a noisy detector of the largest peaks of the input signal. In \cite{dante1,dante2} the nonlinear element is a constant threshold or a neuron's model. In \cite{braunchialvo}  the nonlinear device is a exponential time-dependent double-threshold device, as it is shown in Fig.~\ref{Fig1}


\subsection{Modeling pitch perception}\label{SecPitch}

The understanding of the mechanisms behind pitch perception are also relevant to related issues concerning consonance, music and speech, for example.  Many attempts have been made  to model pitch perception \cite{green,carianisolo,tramo}; however its neural mechanisms are still controversial \cite{ch4,ch5,cariani1,cariani2,ch8,ch9}.  In this context, a simple model  \cite{dante1,dante2}, based on quantifiable and physiologically plausible neural mechanisms was recently proposed to account for key experimental observations related to the missing fundamental illusion and pitch shift.

The assumptions of the model are simple. Let us consider a complex tone $S_c$
formed by adding pure tones of frequencies $f_1=kf_0$, $f_2=(k+1)f_0$, ....,
$f_N=(k+N-1)f_0$ as an input of a nonlinear threshold device. It can be
observed that  the harmonic tone $S_c$ exhibits large amplitude peaks
(asterisks in left panel of Fig.~\ref{Fig1}) spaced at intervals
$T_0=1/f_0$. These peaks are the result of a constructive interference between
the constitutive tones. 

\begin{figure}[htpb]
\begin{center}
\includegraphics[width=4in]{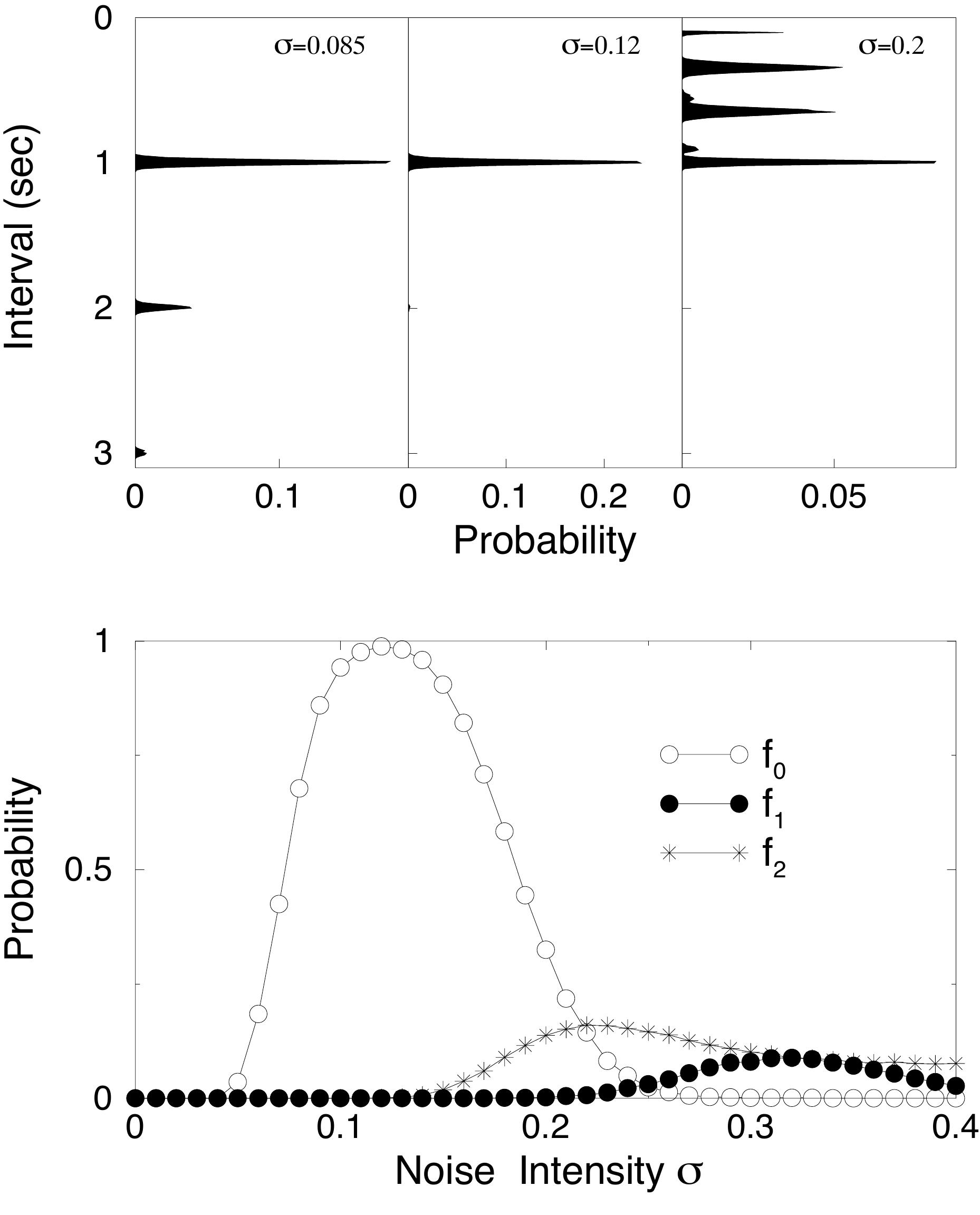}
\end{center}
\caption{\label{ChialvoFig2} Single neuron GSR. Top panels: Density distributions of inter-spike intervals $t$ in the system for three noise intensities. Bottom panel: Signal-to-noise ratio computed as the probability of observing an inter-spike interval of a given $t$ ($\pm 5\%$ tolerance) as a function of noise variance $\sigma$, estimated for the two input signals' time scales: $f_1$ (stars) and $f_2$ (filled circles) as well as for $f_0$ (empty circles). The large resonance is at $f_0$, i.e., a subharmonic which is not present at the input. 
Reprinted with permission from  D.R. Chialvo, O. Calvo, D.L. Gonzalez, O. Piro, and G.V. Savino,  Phys. Rev. E 65, pp. 050902-5(R), (2002). Copyright (2002) from American Physical Society.
} 
 
\end{figure}

The threshold device detects ``statistically" (with the help of noise) the largest peaks of $S_c$, which are spaced by a value corresponding to the fundamental period.
The top panels of Fig.~\ref{ChialvoFig2} show the density distributions of inter-spike intervals
$t$ in the model for three noise intensities. In the bottom panel the
signal-to-noise ratio is computed as the probability of observing an
inter-spike interval of a given $t$ ($\pm 5\%$ tolerance) as a function of
noise variance $\sigma$, estimated for the two input signals' time scales:
$f_1$ (stars) and $f_2$ (filled circles), as well as for $f_0$ (empty
circles). The large resonance occurs at $f_0$, i.e., at a subharmonic frequency which is not present at the input. 

Similar results were obtained using a more elaborated FitzHugh-Nagumo model  \cite{dante1}. The results shown in the bottom panel  of
Fig.~\ref{ChialvoFig2} resemble those described for the stochastic resonance
phenomenon \cite{SR3}. However, in this case, the  optimum
noise intensity for which the system emits the majority of spikes is at a
frequency which is not present in the input. Thus, as happen in the missing fundamental experiments, the model neuron's strongest resonance occurs at a
frequency which is not present in the input.


The results of Fig.~\ref{ChialvoFig2}  are only representative of complex signal composed of harmonic tones, i.e., when frequencies of pure tones composing $S_c$ are integer multiples of the missing fundamental. 
However, the most interesting experimental results in pitch perception are related to the so-called  pitch-shift effect, as was shown in Fig.~\ref{fig_schou}.  
In order to verify if this model is able to reproduce these results, the same threshold device is stimulated by a complex tone $S_c$ formed by adding pure tones of frequencies $f_1=kf_0+\Delta f$, $f_2=(k+1)f_0+\Delta f$, ...., $f_N=(k+N-1)f_0+\Delta f$. In this way, $S_c$ is formed by $N$ equispaced  tones  shifted linearly by $\Delta f$.
Note that if $\Delta f=0$, $f_0$ is the missing fundamental frequency of $S_c$ as in the previous case.

It is expected that the threshold crossings will correspond  preferentially to the highest peaks of $S_c$, produced by the positive interference of the constituent pure tones.  If we think in the simplest case of constructive interference of two tones, it arises from the beating
phenomenon which results in a carrier frequency of $f^+=(f_2+f_1)/2$ modulated
in amplitude with a sinusoidal wave of frequency $f^-=(f_2-f_1)/2$. In this
case, the interval between the highest peaks is equal to the nearest integer
number $n$ of half-periods  of the carrier lying within a half-period of the
modulating signal. For the case of two consecutive higher harmonics of a given
fundamental $f_0$, $f_1=kf_0$ and $f_2=(k+1)f_0$, it can be obtained that
$n=f^+/f^-=2k+1$ and the corresponding interval is $n/f^+=1/f_0 $. If the two
frequencies are linearly shifted in $\Delta f$, it is expected that the most
probable interval between the highest peaks takes place at a rate $f_r$ with $1/f_r=n/f^+=(2k+1)/((2k+1)f_0+2\Delta f)$. 

In the general  case where $S_c$ is composed of $N$ harmonic tones as described above, the expected resonant frequency $f_r$ follows the relation:

\begin{equation}
f_r=f_0+\frac{\Delta f}{k+(N-1)/2}
\label{Deltaf}
\end{equation}

\begin{figure}[htpb]
\begin{center}
\includegraphics[width=5in]{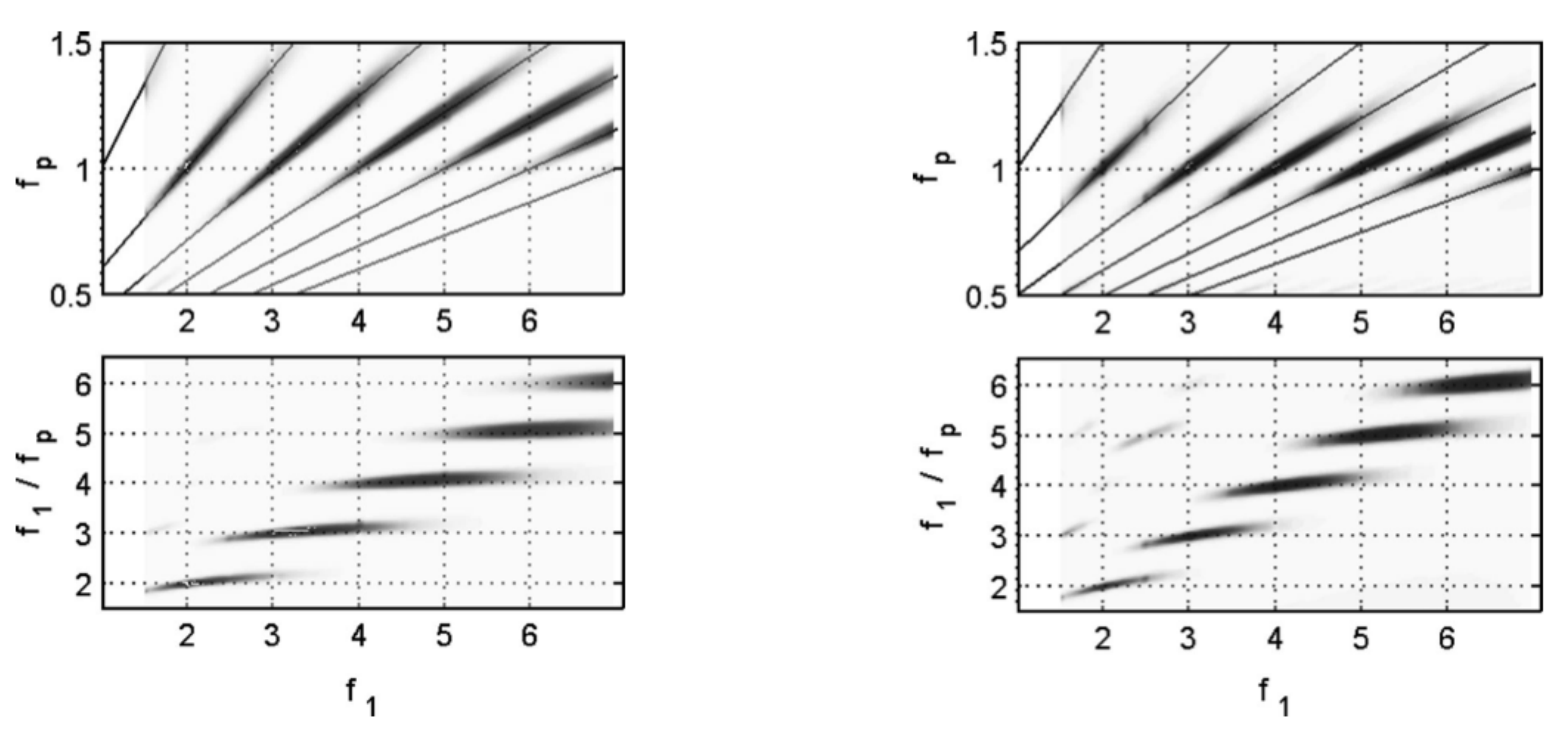}
\end{center}
\caption{\label{ChialvoFig3y4PRE} Pitch shift simulations for $N=2$ (left panels) and $N=3$ (right panels) frequency signals. Top: The probability (as gray scale) of observing a spike with a given instantaneous firing frequency $f_p$ (in the ordinate) as a function of the frequency $f_1$ of the lowest of two components of the input signal (abscissa). The family of lines is the theoretical expectation (Eq.(\ref{Deltaf})) for $N=2,3$ and $k=1-7$. Bottom: The same data from the top panels are replotted as input-output frequency ratio vs input frequency $f_1$ ($f_0=1$Hz) ($f_p$ corresponds to $f_r$ in the notation of Eq.\ref{Deltaf}). Reprinted with permission from  D.R. Chialvo, O. Calvo, D.L. Gonzalez, O. Piro, and G.V. Savino,  Phys. Rev. E 65, pp. 050902-5(R), (2002). Copyright (2002) from American Physical Society.} 
 
\end{figure}

The agreement between simulations and the theory is remarkable as demonstrated by the results in Fig.~\ref{ChialvoFig3y4PRE} where the simulations and the theoretical lines matches perfectly.


\begin{figure}

\begin{center}
\includegraphics[width=6in]{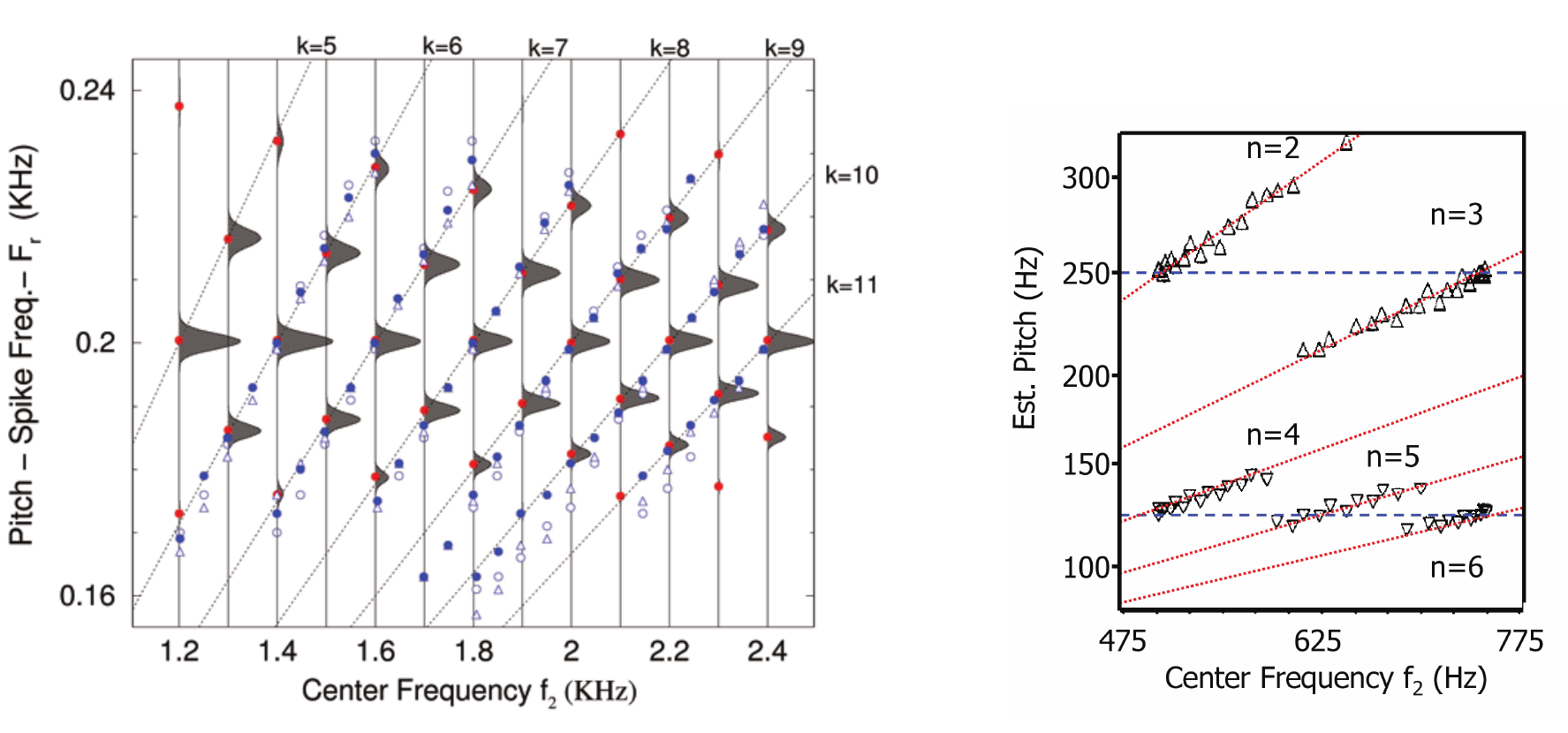}
\end{center}
\caption{\label{ChialvoFig4} Simulations and experiments of pitch shift. The
  left panel shows that the  theoretical prediction of Eq.\ref{Deltaf}
  superimposes exactly with the result of simulations (Grey Histograms,
  $k=5-9$ and $N=3$) and reported pitch from Schouten's experiments (circle
  and triangle symbols) \cite{schouten}. Right  panel (b): The theoretical expectation from Eq.(\ref{Deltaf}) is superimposed here onto the experimental results of Cariani and Delgutte \cite{cariani2}. The physiological pitches estimated from the highest peak of the interspike interval distribution in response to two variable ($500-750$ Hz) carrier AM tones with modulation frequencies (horizontal dashed lines) of 125 Hz (downward triangles) or 250 Hz (upward triangles) fit the predictions of Eq.\ref{Deltaf} (diagonal dotted lines with $N=3$).
The agreement in both cases is parameter independent. 
Reprinted with permission from  D.R. Chialvo, Chaos 13, pp. 1226-30 (2003). Copyright (2003) by the American Institute of Physics.} 
\end{figure}

If $f_0=200$Hz, $N=3$ and $k=6$, the stimulus $S_c$ has the same features as those used in \cite{schouten}.  Panel (a) of Fig.~\ref{ChialvoFig4} shows the results of these simulations superimposed with psycho-acoustical pitch reports from Schouten et al's experiments \cite{schouten}. 
The results of the simulations are presented as histograms of inter-spike intervals produced by the neuron model. The experimental results report the pitch detected by the subjects in the experiments.  
Panel (b) of Fig.~\ref{ChialvoFig4} shows the excellent agreement between Eq.(\ref{Deltaf}),  simulation data and the  pitch estimated from the predominant inter-spike interval in the discharge patterns of cat auditory nerve fibers in response to complex tones \cite{cariani1,cariani2}. It should be noticed that, in both cases, this agreement  is parameter independent.


The simplicity of the GSR model discussed until now contrasts with previous ideas which suggested complicated
mechanisms mediated by relatively sophisticated structures not yet identified as timing nets \cite{carianisolo}, delay lines \cite{ch4}, neural networks \cite{ch9}, oscillators in combination with integration circuits \cite{ch22},   and others \cite{ch23}. Finally, it is important to remark that these results were replicated by recent numerical experiments on a detailed model of the cochlea \cite{coclea2010}.

\subsection{Modeling Dansgaard-Oeschger events}\label{sec:D-O}

As discussed in section \ref{IntroGlacial}, many of the Dansgaard-Oeschger temperature shifts (see Fig.~\ref{FigO18}) are spaced by intervals (or integer multiples) of about $1470$ years, but the origin of this apparent regularity is still a matter of ongoing debate.
One of the most prolific approaches to understand the DO events was done recently in terms of Ghost Stochastic Resonance \cite{glacial}. In what follows we will show two models which, at different scales and based on GSR,
aim to explain the recurrence properties of these warming events. 

\subsubsection{Ocean-Atmosphere  Model}

It has been hypothesized \cite{glacial} that the $1470$-year recurrence time of the DO events could result from the presence of two centennial solar cycles, the DeVries and Gleissberg cycles with leading spectral components corresponding to periods near 210 and 87 years \cite{braun7a,braun7b,braun8,braun10,braun11} respectively. 
A carefully inspection of these frequency cycles shows that they are approximately the 7 and 17 harmonic superior of 1470, which leads to the hypothesis that DO events could
be caused by a GSR mechanism.  In order to test this hypothesis, a coupled ocean-atmosphere model (CLIMBER-2)
\cite{glacial} was forced with the two mentioned solar frequencies: 

\begin{equation} 
F(t)=-A_1\cos(2\pi f_1 t +\phi_1)-A_2\cos(2\pi f_2 t +\phi_2)+K
\label{eqBraunNature}
\end{equation}

For simplicity, this forcing was introduced as a variation in freshwater input with
$f_1=1/210$~years$^{-1}$  and $f_2=1/86.5$~years$^{-1}$.  The motivation for
using a freshwater forcing is as follows: Many ocean-atmosphere models indicate that
changes in the freshwater flux to the area of deep buoyancy convection in the 
northern North Atlantic could trigger shifts between different modes of the 
thermohaline (i.e., density-gradient driven) ocean circulation, since the
density of ocean water depends on salinity. During glacial times,  solar
forcing is expected to have a notable influence on the freshwater budget in
the northern North Atlantic, for example due to solar-induced variations in
the mass budget of the surrounding continental ice sheets.  

The amplitudes $A_1$ and $A_2$ as well as the phases $\phi_1$ and $\phi_2$ are parameters of the
forcing. The constant $k$ mimics changes in the background climate compared
with the Last Glacial Maximum, which is considered as the underlying climate
state.  Further details of the simulations can be found in \cite{glacial}.
Within a wide range of forcing parameters, this kind of perturbation to the
model produces events similar to the DO ones. The simulated events represent
transitions between a stadial (``cold") and interstadial
(``warm") mode of the North Atlantic thermohaline ocean circulation. 

\begin{figure}[htpb]
 
\begin{center}
\includegraphics[width=4in]{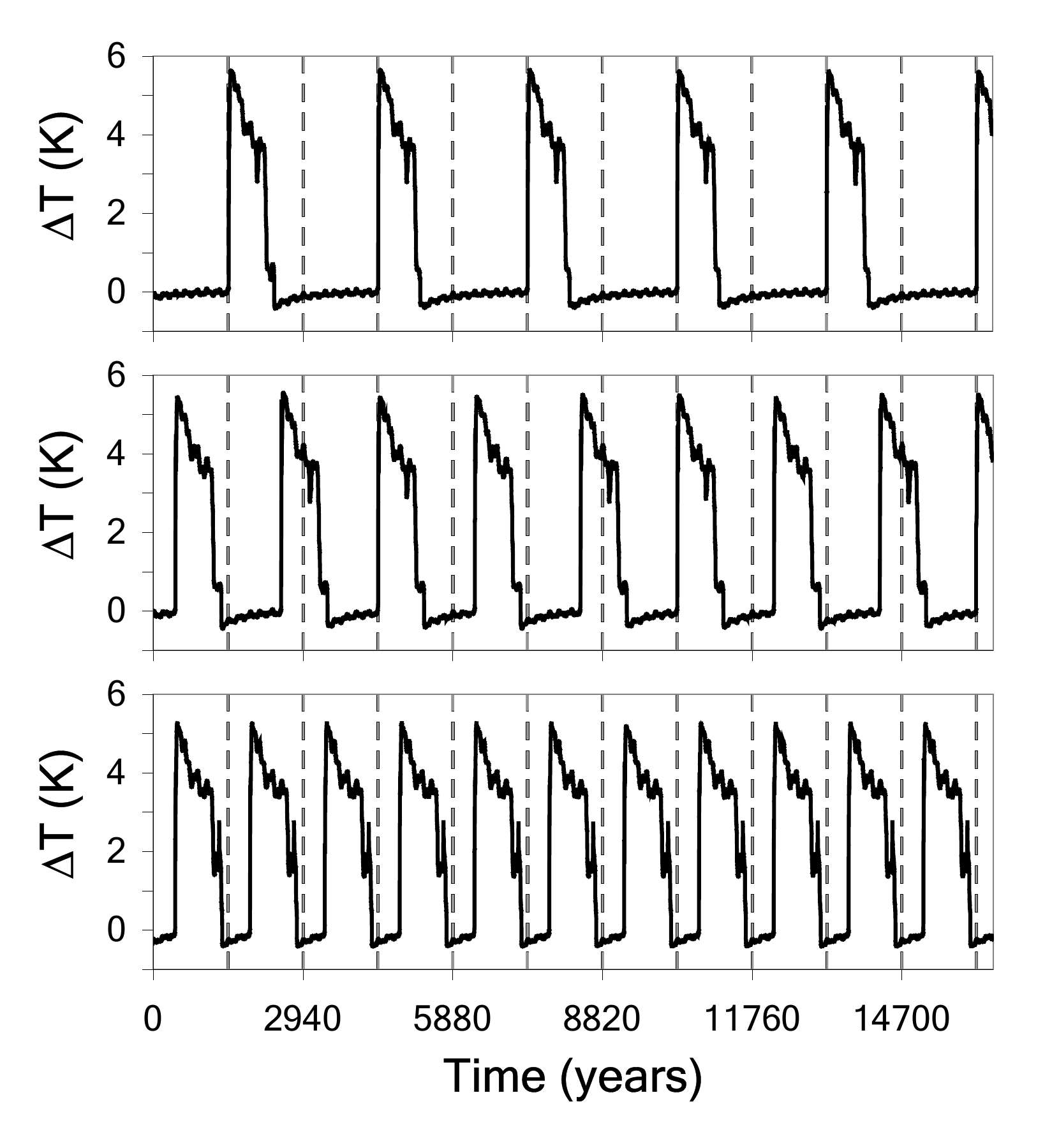}
\end{center}
\caption{\label{Fig2BraunNature} $1470$ years recurrence time scale from the
  CLIMBER-2 model simulations. All three panels show changes $\Delta T$  in
  Greenland surface air temperature according to simulations of Braun et
  al. \cite{glacial}. The amplitudes of the forcing (Eq.\ref{eqBraunNature})
  are $A_1=A_2=10mSv$.  Top, middle and bottom panels correspond to three different   values of $K$ ($K=-9$~mSV, $-14$~mSv  and $-19$~mSv) respectively. The dashed lines indicate the position of the global minima in the forcing and are spaced by   $1470$~years. We can observe that in all cases the events repeat strictly  with a period of either $1470$~years (bottom) or integer   multiples of that value (top, middle). Note that despite the period of   four times $1470$~years in the middle panel, the average inter-event spacing is $1960$~years, and all events in that panel can be divided into three groups, each of which  contains only events which recur exactly in integer multiples of   $1470$~years. Reproduced from 
   \cite{glacial}.}  
 
\end{figure}

In the response of the model three  different regimes exist: a ``cold regime"
in which the thermohaline ocean circulation persists in the stadial mode, a ``warm
regime'' in which the interstadial mode is stable and a ``Dansgaard-Oeschger
regime" in which cyclic transitions between both modes occur. These transitions
result in abrupt warm events in the North Atlantic region, similar to the DO
events as shown in the results of Fig.~\ref{Fig2BraunNature}. 


Events spaced by $1470$~years are found within a continuous range of
forcing-parameter values. Indeed, this timescale is robust when the phases, the
amplitudes and even the frequencies of the two forcing cycles are changed over some
range (see \cite{glacial}).  Noise, when added to the periodic
forcing, is unable to affect the preferred tendency of the events to recur
almost exactly in integer multiples of $1470$~years.

These simulations, which reproduce some of the characteristic recurrence
properties of the DO events in the paleoclimatic records, clearly show that a
ghost stochastic resonance is at work in the model simulations, making GSR a potential candidate to explain the $1470$~years recurrence time scale of Dansgaard-Oeschger events during the last ice age.

\subsubsection{Conceptual DO Model}

As happened with the analysis of pitch perception in section \ref{SecPitch}, a low dimensional modeling approach can be followed in order to isolate and understand the main dynamical mechanism present in the behavior of the complex model analyzed in last section.

The stability of the simulated $1470$~year climate cycle turns out to be a
consequence of two well-known properties of the thermohaline circulation: its
long characteristic timescale, and the high degree of nonlinearity (that is,
the threshold character) inherent in the transitions between the two simulated
modes of the thermohaline circulation. A very simple conceptual model that
only incorporates these two properties is able to mimic key features of
CLIMBER-2, i.e., the existence of three different regimes in the model
response, the frequency conversion between forcing and response (that is, the
excitation of millennial-scale spectral components in the model response that
do not exist in the forcing) and the amplitude dependence of the period in the
model response \cite{braunchialvo}. The general idea of this model is that DO events represent highly nonlinear switches between two different climate states corresponding to the stadial (``cold") and interstadial (``warm") modes of the glacial thermohaline circulation. 

This conceptual model is based in three key assumptions:

\begin{enumerate}
\item DO events represent repeated transitions between two different climate states, corresponding to warm and cold conditions in the North Atlantic region.
\item These transitions are rapid compared to the characteristic life-time of
  the two climate states and take place each time a certain threshold is crossed.
\item With the transition between the two states the threshold overshoots and afterwards approaches equilibrium following a millennial-scale relaxation process. 
\end{enumerate}
These three assumptions, which are supported by high-resolution paleoclimatic records and/or by simulations with a climate model  \cite{braunchialvo}  can be implemented in the following way: A discrete index $s(t)$ that indicates the state of the system at time $t$ (in years) is defined. The two states, warm and cold, correspond with the values $s=1$ and $s=0$ respectively. A threshold function $T(t)$ describing the stability of the system at time $t$ is also defined.

The next step is to define the rules for the time evolution of the threshold
function $T(t)$. When the system shifts its state, it is assumed that a
discontinuity exists in the threshold function: With the switch from the warm state to the cold one (at $t=t'$ in panel (d) of Fig.~\ref{Fig1}) $T$ takes the value $A_0$. Likewise, with the switch from the cold state to the  warm one (at $t=t''$ in panel (d) of Fig.~\ref{Fig1})  $T$ takes the value $A_1$. As long as the system does not change its state, the evolution  of $T$ is assumed to be given by a relaxation process
\begin{equation}
T(t)=(A_s-B_s).\exp(-\frac{t-\delta_s}{\tau_s})+B_s,
\end{equation}

where the index $s$ stands for the current state of the model (i.e., $s=1$ is
the warm state and $s=0$ the cold one), $\delta_0$ labels the time of the last switch from the warm state into the cold one and $\delta_1$ labels the time of the last switch from the cold state into the warm one.
The third assumption is that the change from one state to another happens when
a given forcing function $f(t)$ crosses the threshold function. 
In the right panel of Fig.~\ref{Fig1} we plot an schematic representation of the
threshold function when the model is forced with a bi-sinusoidal
input with frequencies $f_1$ and $f_2$ which are the second and third harmonic
of a given $f_0$. As we have seen in the previous sections, the peaks of the forcing repeat with a period of $1/f_0$~years, despite the absence of this period in the two forcing series. 

A stochastic component represented by white noise of zero mean and amplitude $D$ is added in order to take non-periodic forcing components into account. 
 
\begin{figure}[htpb]

\begin{center}
\includegraphics[width=6in]{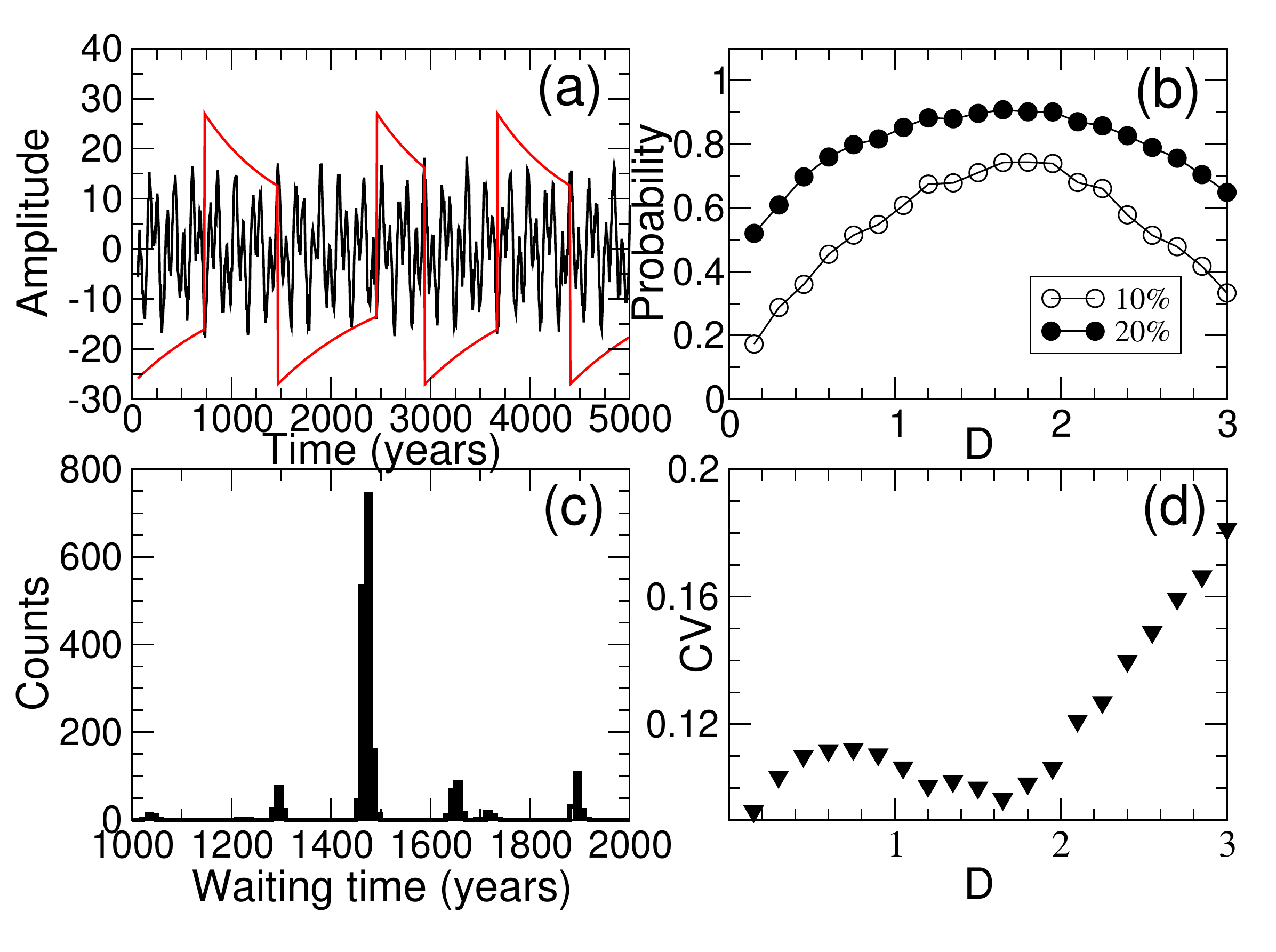}
\end{center}
\caption{\label{Fig2ConcpMod} Conceptual model for DO events. Panel (a):
  Forcing signal plus optimal noise and threshold function $T(t)$. Panel (b):
  Signal to noise ratio calculated as that fraction of inter-event waiting
  times that has values within $10\%$ and $20\%$ of $1470$~years
  (right). Panel (c):  Histograms of the inter-event waiting times of the
  simulated events  for optimal noise (left). Panel (d): Coefficient of
  variation as a function of the noise amplitude.  Its minimum value indicates
  the resonance.  The parameters are: $A_0=-27$, $A_1=27$, $B_0=3$, $B_1=-3$, $\tau_0=1200$, $\tau_1=800$ } 

\end{figure}

In Fig.~\ref{Fig2ConcpMod}  the response of the model is shown for different
values of noise amplitudes.  It can be observed that for an optimal noise
amplitude the waiting time distribution of the simulated events is peaked at
$1470$~years. This is reflected in the minimum of the  coefficient of
variation of the waiting times (i.e.,  the standard deviation divided by its
average) and in the maximum of the signal to noise ratio (calculated as that
fraction of inter-event waiting times that has values around $1470$~years) for
the optimal values of noise.


These results demonstrate that a low-dimensional model, constructed from the dynamics of the
events manifested in the much more complex ocean-atmosphere model
CLIMBER-2 is able to exhibit GSR. 
For completeness, it need to be discussed how  variations in any one of the input frequencies could alter the results, in other words how stable the GSR phenomenon is under these changes. This analysis was addressed in Braun et al. \cite{glacial} who reported in their supplementary material that such changes typically alter the average inter-event waiting time by less than 20 per cent, consistent with the results presented in the supplementary material of a successive study \cite{braunchialvo}. Another effect discovered in both studies is that the dispersion of the inter-event waiting time distribution increases when the system is driven out of the resonance case by a change in any of the forcing frequencies. Thus, the phenomenon was found to be stable under changes of the input frequencies.

\section{Related work}

\subsection{Binaural pitch perception}\label{SecBinau}

Besides the question of {\it how} pitch is perceived, another contested debate relates to {\it where} perception
takes place. Although interval statistics of the neuronal firings \cite{cariani1,cariani2} show that pitch
information is already encoded in peripheral neurons, under other conditions pitch perception can take place at a higher level of neuronal processing \cite{pantev}. A typical example is found in binaural experiments, in which the two components of a harmonic complex signal enter through different ears. It is known that in that case a (rather weak) low-frequency pitch is perceived. This is called ``dichotic pitch'', and can also arise from the binaural interaction between broad-band noises. For example, Cramer and Huggins \cite{cramer} studied the effect of a dichotic white noise when applying a progressive phase shift across a narrowband of frequencies, centered on $600$ Hz, to only one of the channels. With monaural presentation listeners only perceived noise, whereas when using binaural presentation over headphones, listeners perceived a $600$Hz tone against a background noise.

Recent work shows that the binaural effects described above can be explained in GSR terms. The model comprises a three neurons structure. Two of them receive one single component of the complex signal, so that each neuron represents detection at a different auditory channel in a binaural presentation,  acting upon a third processing neuron \cite{balenzuela05}. These results showed that the higher-level neuron is  able to perceive the pitch, hence providing a neural mechanism for the binaural experiments. The membrane potential of the neurons were simulated via a Morris-Lecar model \cite{morlec}. The two input neurons were unidirectionally coupled to the response neuron via a synaptic coupling model \cite{destexhe}. Details of the modeling can be found in \cite{balenzuela05}.  

\begin{figure}

 \begin{center}
\includegraphics[width=6in]{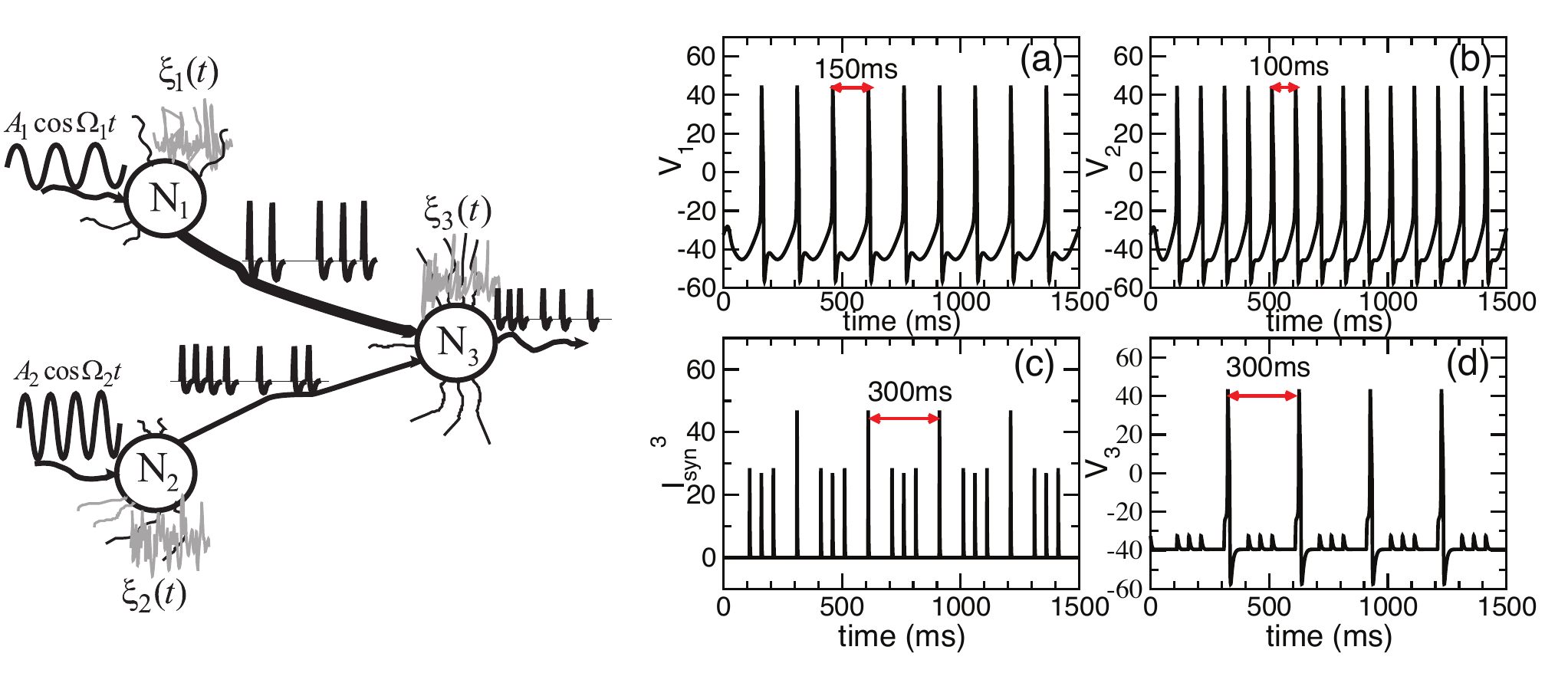}
\end{center}
\caption{Binaural GSR. Left panel: Schematic representation of the binaural
 scenario. Right Panel: Deterministic responses to a  binaural harmonic signal. 
 The membrane potential for the three neurons is shown: (a,b)  input neurons, (d) processing output neuron. The synaptic current acting on the output neuron is shown in plot (c). The two inputs neurons are stimulated with two sinusoidal signals. Reprinted with permission from P. Balenzuela and J. Garc\'ia-Ojalvo,, Chaos 15, pp. 023903-10, (2005). Copyright (2005) from American Institute of Physics and  Y. V. Ushakov, A. Dubkov and B. Spagnolo,  Phys. Rev. E  81, pp. 041911-23 (2010). Copyright (2010) from American Physical Society. } 
\label{FigBinaural1}

\end{figure}

The right panels of Fig.~\ref{FigBinaural1} show the different stages of the
simulation when input neurons are stimulated by pure tones with periods of
$150$~ms and $100$~ms respectively. The panels (a) and (b) display the
membrane potential with spikes exhibiting the same period of input signals.
Panel (c) shows the synaptic current elicited by these two neurons onto the
processing output neuron. We can observe that the maxima of this
current are spaced at intervals of $300$~ms, which correspond to the missing fundamental
frequency. The same intervals appear in the spike trains of the processing
neuron as seen in panel (d). 


\begin{figure}[htpb]
 
\begin{center}
\includegraphics[width=5in]{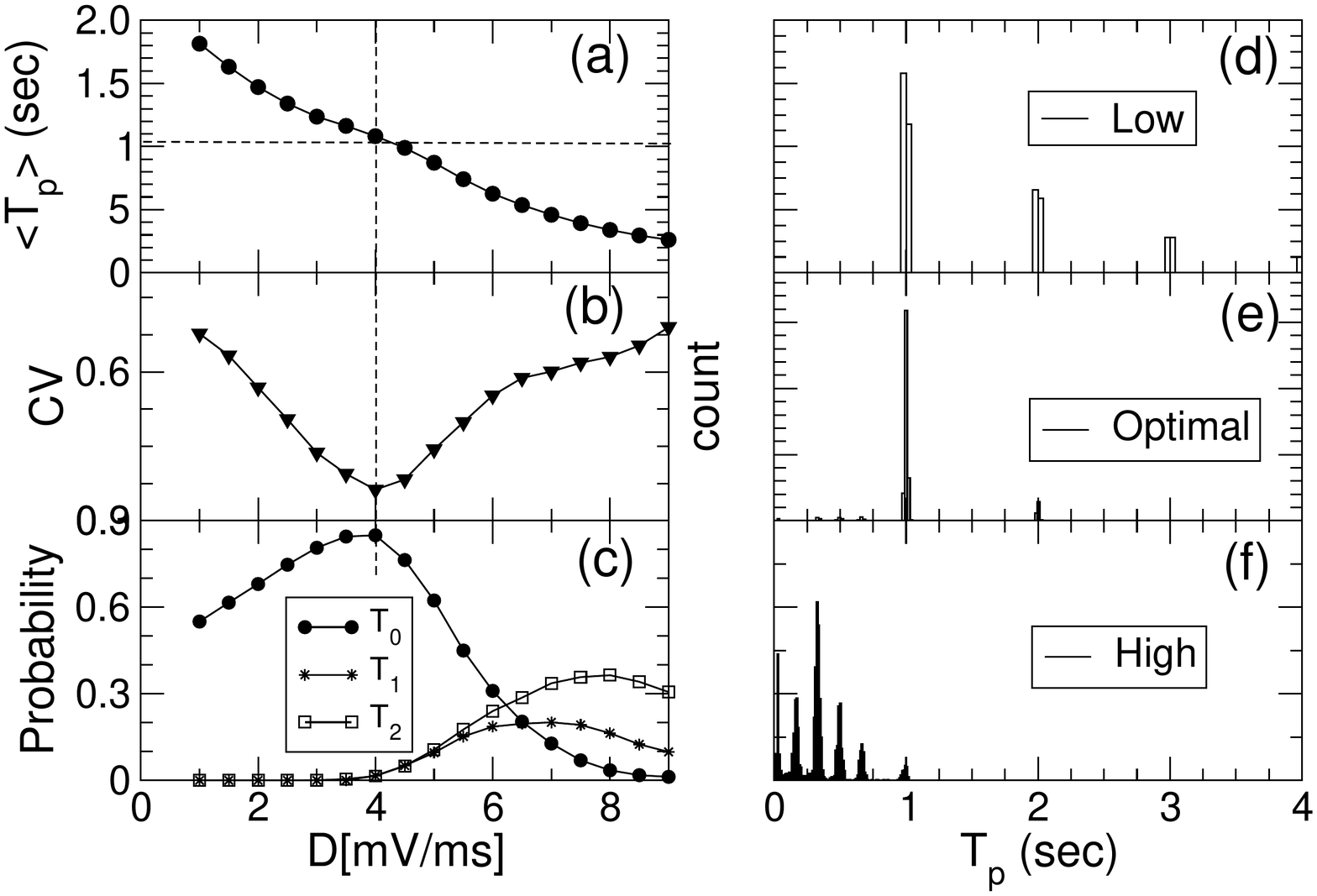}
\end{center}

\caption{\label{FigBinaural2}  Binaural GSR. Left panels: response of the processing neuron as a function of the noise amplitude: (a)
mean time between spikes $\langle T_p\rangle $, (b)  coefficient of variation, CV,  and (c) signal to noise ratio measured as the fraction of pulses spaced around $T_0=1/f_0$, $T_1=1/f_1$ and $T_2=1/f_2$, as a function of the noise amplitude in the processing neuron, $D$. Right panels: probability distribution functions of the
time between spikes $T_p$ for three values of the noise amplitude $D$: (d) low (e) optimal,  and (f) high values. Reprinted with permission from P. Balenzuela and J. Garc\'ia-Ojalvo,, Chaos 15, pp. 023903-10, (2005). Copyright (2005) from American Institute of Physics.  } 
 
\end{figure}

In biological neural networks, each neuron is connected to thousands of
neurons whose synaptic activity could be represented as ``synaptic
noise". In this configuration this effect is taken into account by adding a
white noise term of zero mean and amplitude $D_i$ in the input neurons
($i=1,2$) and of amplitude $D$ in the processing neuron's membrane
potential. The firing process of this neuron is then governed by noise. In
the left panels of Fig.~\ref{FigBinaural2} we plot the mean time between
spikes $\langle T_p\rangle $ (panel (a)), the coefficient of variation
($CV=\sigma_p/\langle T_p\rangle $, panel (b)) and the signal to noise ratio
measured as the fraction of pulses spaced around $T_0=1/f_0$, $T_1=1/f_1$ and
$T_2=1/f_2$ (panel (c)) as a function of the noise amplitude in the processing
neuron, $D$. Right panels show the probability distribution functions of the
time between spikes $T_p$ for three values of the noise amplitude $D$: (d) Low
(e) optimal and (f) high values. Note the remarkable agreement between these
results and those obtained for a single neuron in Fig.~\ref{ChialvoFig2},
which confirm the robustness of GSR.

\begin{figure}[htpb]
\begin{center}
\includegraphics[width=4in]{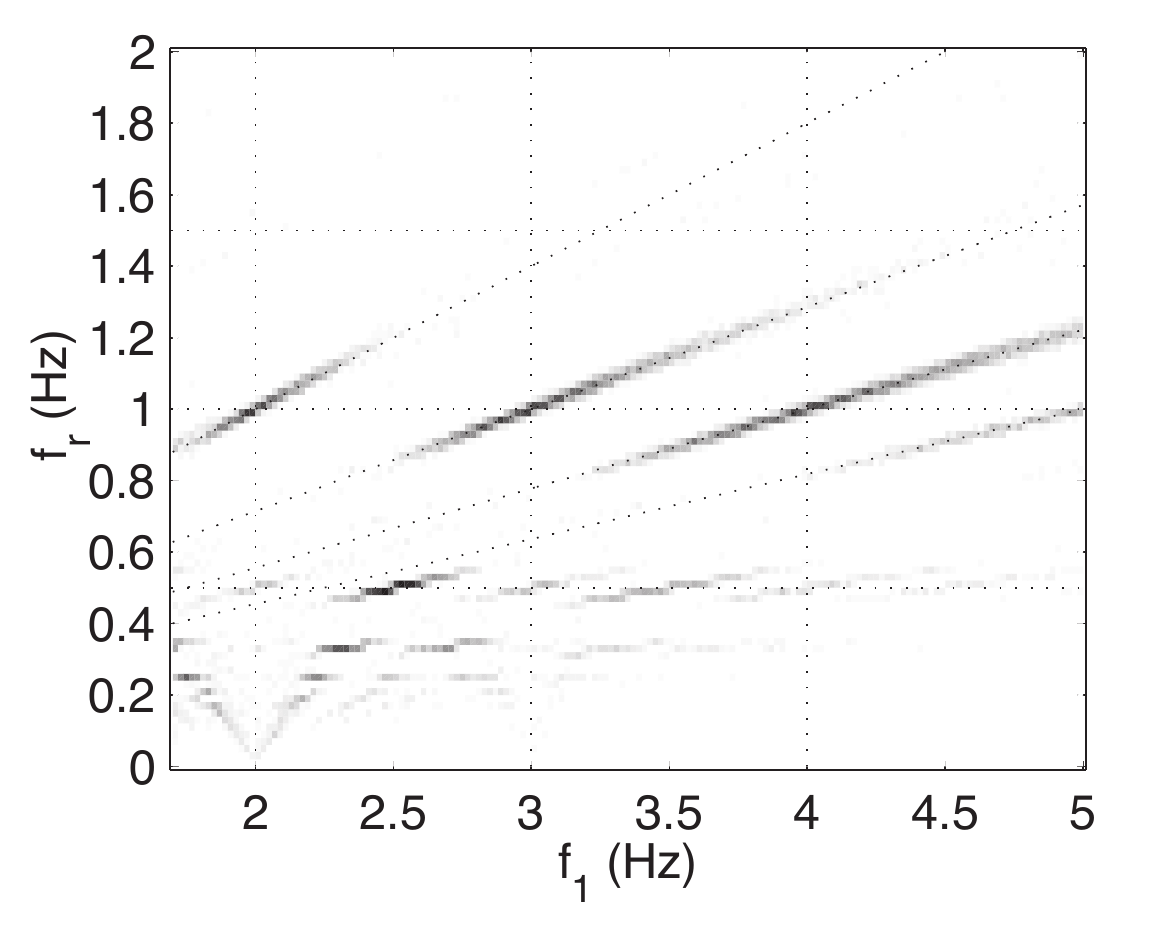}
\end{center}
\caption{\label{FigBalenInHarm} Binaural pitch shift simulations. Probability
  of observing a spike in the processing neuron with instantaneous rate $f_r$
  (in gray scale) as a function of the frequency $f_1$ of one of the input
  neurons. Note the remarkable agreement of the responses following the lines
  predicted by Eq. \ref{Deltaf} for k=2,3,4,5 (dashed lines from top to
  bottom). Reprinted with permission from P. Balenzuela and J. Garc\'ia-Ojalvo,, Chaos 15, pp. 023903-10, (2005). Copyright (2005) from American Institute of Physics. } 
 
\end{figure}

The next step was to check if  this model  reproduces the pitch shift experiments sketched in Fig.~\ref{fig_schou} and follows the theoretical predictions of  Eq.\ref{Deltaf}. 
The pure tones driving the input neurons have frequencies $f_1=kf_0+\Delta f$ and $f_2=(k+1)f_0+\Delta f$. 
Fig.~\ref{FigBalenInHarm} shows the results of the simulations. The instantaneous frequency $f_r=1/T_p$  follows the straight lines predicted in Eq. \ref{Deltaf} for $N=2$ , $k=2-5$ and $f_0=1$Hz. 
It is important to notice that even though the linear superposition of inputs is replaced by a coincidence detection of spikes  in this configuration,   the preferred frequencies response of the output neuron follows the theoretical predictions made for linear interference between tones. This is consistent with the arguments used to deduce 
Eq. \ref{Deltaf}, which looks for the coincidence of maxima  of the harmonic tones. Here, the output neuron detects coincidence of spiking neurons, which also takes place preferentially at the maxima of harmonic inputs.


 A brain structure candidate for the dynamics of the processing neuron is the inferior colliculus, which receives multiple inputs from a host or more peripheral auditory nuclei. Details of the physiology of this nuclei are still uncertain, but enough evidence suggests that temporal and frequency representation of the inputs are present in the spike timing of their neurons. The results in this section suggest that the neurons in this nuclei can exhibit the dynamics described here, thus participating in the perception of the binaural pitch. The main consequence of these observations is that pitch information can be extracted mono or binaurally via the same basic principle, i.e., ghost stochastic resonance, operating either at the periphery or at higher sensory levels.
 
 A similar three neurons arrangement was studied in  \cite{giraudo}, where two
 input neurons act on  a third processing neuron by means of excitatory
 connections. The input neurons were described as modification of the
 Ornstein-Uhlenbeck diffusion process \cite{OU},  where periodic signals of frequencies $f_1=2f_0+\Delta f$ and $f_2=3f_0+\Delta f$ were added to the drift coefficients. The processing neuron was also simulated via a Ornstein-Uhlenbeck diffusion process.   Simulations in this model  confirmed the GSR phenomena in both the harmonic ($\Delta f=0$) and the inharmonic ($\Delta f \ne 0$) cases  \cite{giraudo}.

\subsection{Beyond pitch: Consonance and Dissonance}

The elucidation of the mechanisms intervening in the perception and processing of complex signals in auditory system  is relevant beyond the identification of pitch. An open challenge in this field is to understand the physiological basis for the phenomena of consonance and dissonance \cite{ConDis10,ConDis13,ConDis14}. Consonance is usually   referred to as the pleasant  sound sensation produced by certain combinations of two frequencies played simultaneously. On the other hand, dissonance is the unpleasant  sound heard with other frequencies combinations \cite{ConDis15}. The oldest theory of consonance and dissonance is due to Pythagoras, who observed that the simpler the ratio between two tones, the more consonant they will be perceived. For example, the consonant octave is characterized by a $1/2$ frequency ratio between two tones, meanwhile the dissonant semitone is characterized by a $15/16$ ratio. Helmholtz \cite{helmholtz} analyzed this phenomenon in the more sophisticated  scenario of complex tones. When two complex tones are played together, it happens that for some combinations (simple ratio $n/m$) the harmonic frequencies match, while in other cases (more complicated ratios $n/m$) they do not. As the frequency ratio becomes more {\it complicated}, the two tones share fewer common harmonics leading to an unpleasant beating sensation or dissonance.  

Ushakov et al.  \cite{ConDis} used a neuronal configuration similar to the one described in the left panel of Fig.~\ref{FigBinaural1} to investigate the phenomenon of consonance and dissonance in tonal music.  In this configuration, a complex input composed of two harmonic signals (with frequencies $f_1$ and $f_2$) is transformed by this simple  sensory model into different types of spike trains, depending of the ratio of the inputs frequencies. More regular patterns in the inter-spike interval distribution (ISI)  were associated to consonant accords whereas less regular spike trains and broader ISI distributions corresponded to dissonant accords.


\begin{figure}[htpb]

\begin{center}
\includegraphics[width=5in]{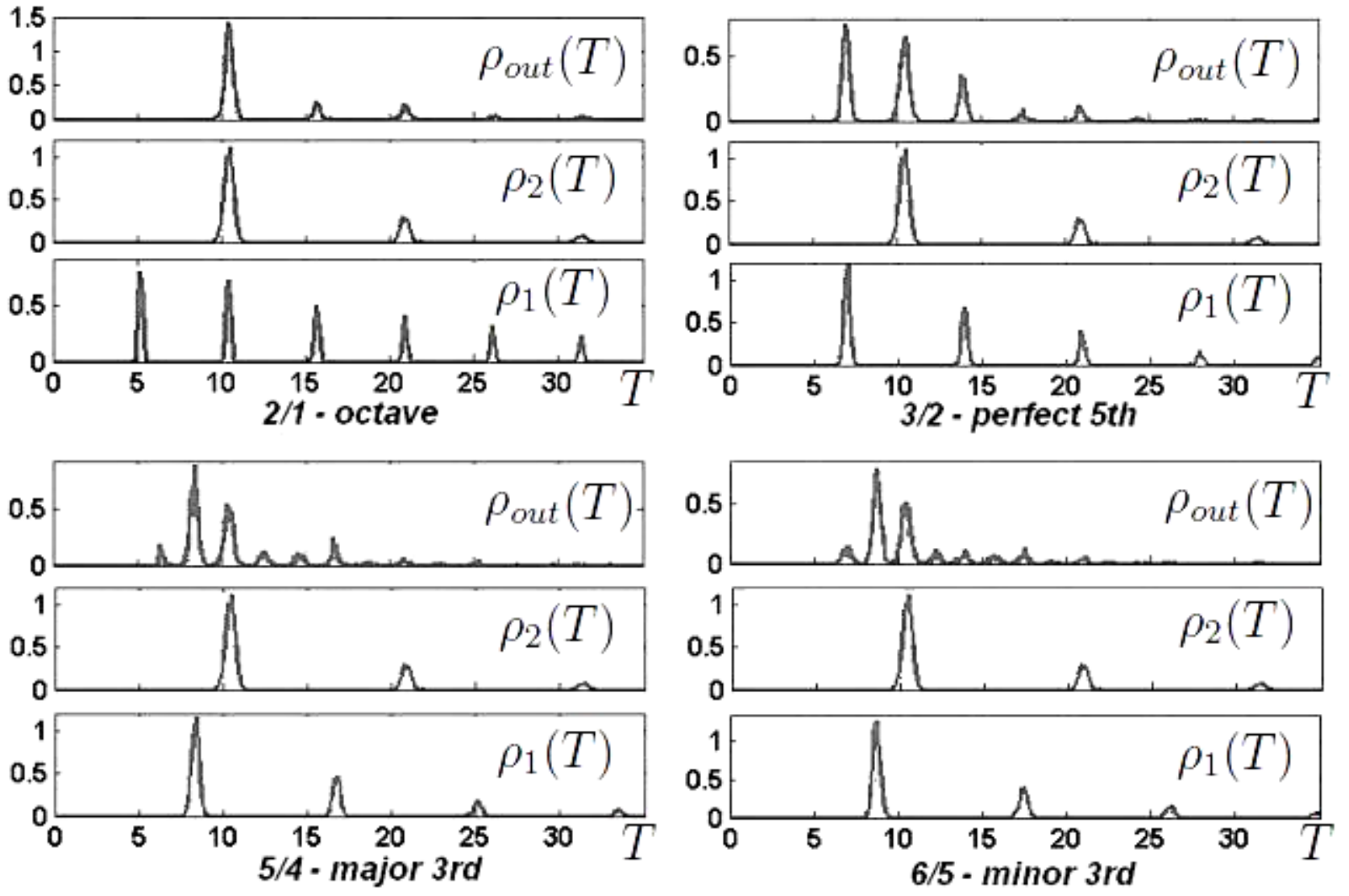}
\end{center}
\caption{\label{Fig5ConsDis} Inter-spike interval distributions of the consonant accords: octave (2/1), perfect fifth (3/2), major third (5/4), and minor third (6/5) for both input neurons ($\rho_1$ and $\rho_2$) and the processing neuron ($\rho_{out}$). Under each picture there is the ratio of frequencies ($m/n$) as well as the common musical terminology. Reprinted with permission from  Y. V. Ushakov, A. Dubkov and B. Spagnolo,  Phys. Rev. E  81, pp. 041911-23 (2010). Copyright (2010) from American Physical Society.} 
 
\end{figure}

\begin{figure}[htpb]
\begin{center}
\includegraphics[width=5in]{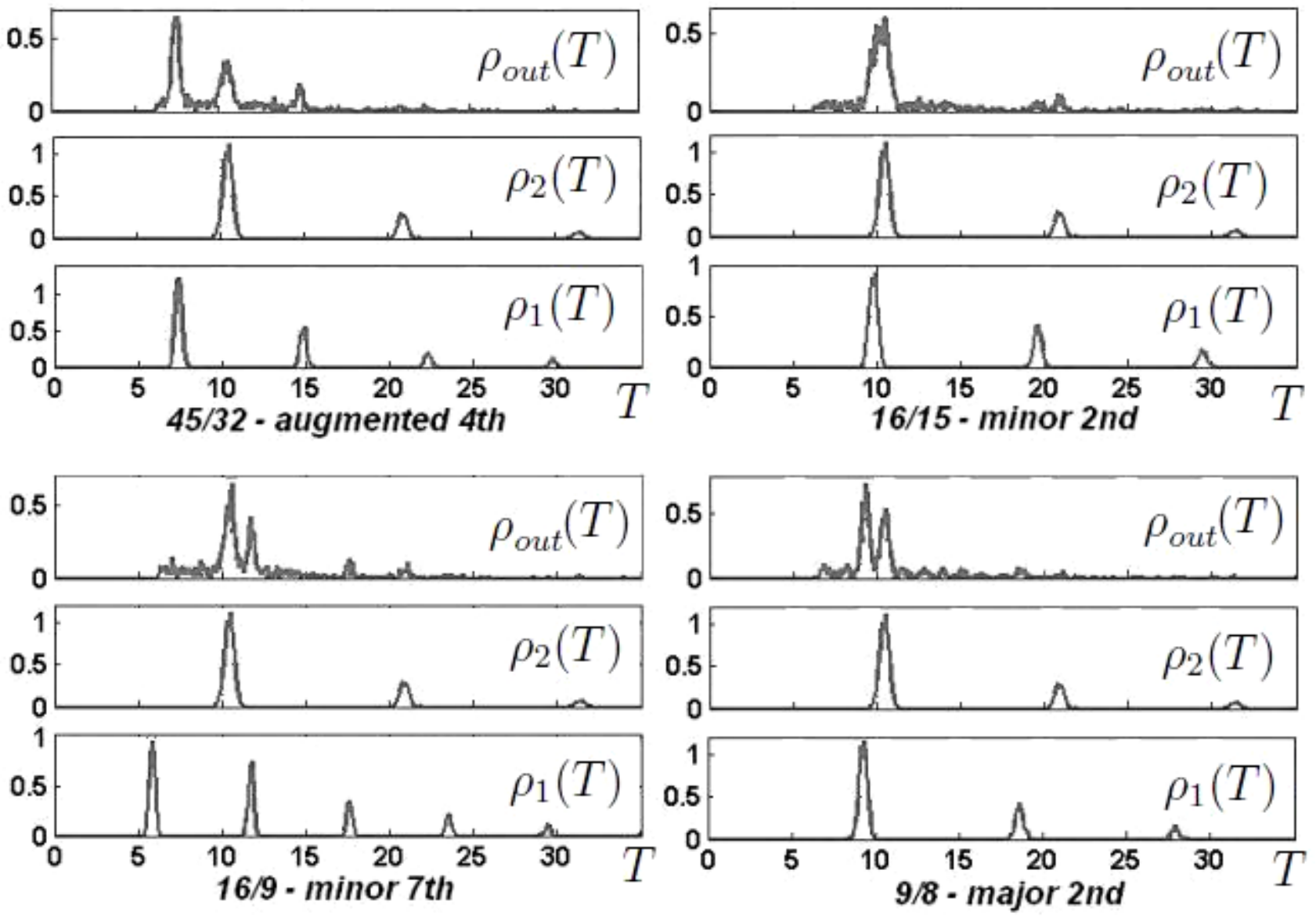}
\end{center}
\caption{\label{Fig6ConsDis} Inter-spike interval distributions of the dissonant accords: major second (9/8), minor seventh (16/9), minor second (16/15), and augmented fourth (45/32). Under each picture there is the ratio of frequencies (m/n) as well as the common musical terminology.  Reprinted with permission from  Y. V. Ushakov, A. Dubkov and B. Spagnolo,  Phys. Rev. E  81, pp. 041911-23 (2010). Copyright (2010) from American Physical Society.} 
 
\end{figure}

Fig.~\ref{Fig5ConsDis} shows the ISI distributions of the processing neuron for a group of consonant accords: an octave ($2/1$), a perfect fifth ($3/2$), a major third ($5/4$) and a minor third ($6/5$). Notice the peaks in the distribution  $\rho_{out}$ which are not present in the input patterns of $\rho_1$ and $\rho_2$. The results for dissonant accords are shown in Fig. ~\ref{Fig6ConsDis}. Further work of the same authors \cite{yuri2011} found that consonance can be estimated by the entropy of the ISI distributions, being smaller for consonant inputs. These results suggest that the entropy of the neural spike trains is an objective quantifier of this very subjective percept. 


\subsection{A dynamical model for Dansgaard-Oeschger events}

In section \ref{sec:D-O}  two models were presented in order to explain the
characteristic $1470$ years observed recurrence time of DO events in
terms of a bi-sinusoidal forcing with frequencies  close to the main
spectral components of two reported solar cycles, the
DeVries/Suess and Gleissberg cycles. In this section a model \cite{sebas} for
the evolution of Greenland paleo-temperature during the last about $80,000$ years
(an interval that comprises the last ice age as well as the current warm
age called Holocene) is described. 


\begin{figure}[htpb]
 
\begin{center}
\includegraphics[width=4in]{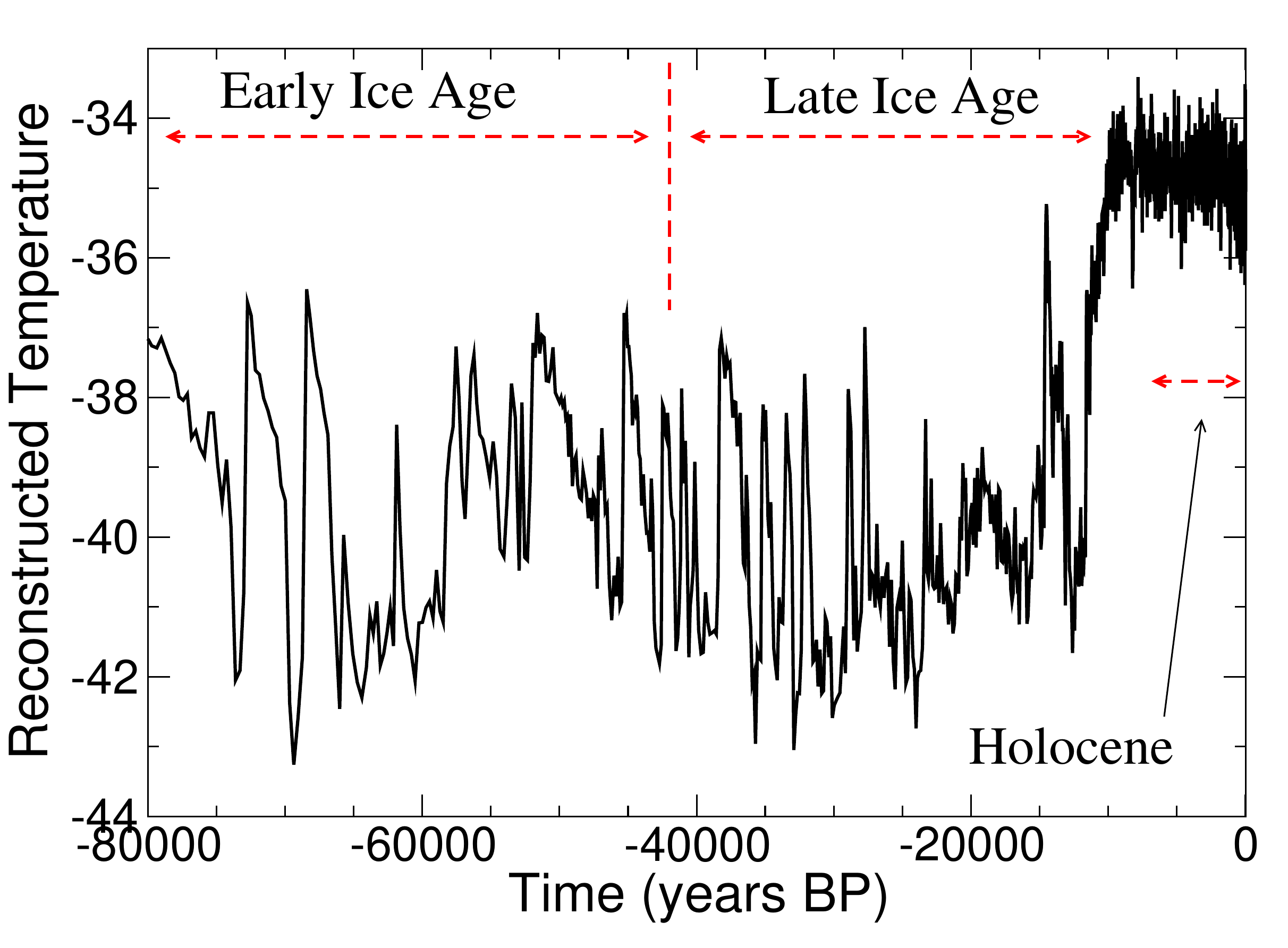}
\end{center}
\caption{\label{Fig1Sebas} Reconstructed Greenland temperature based on the
  ratio of two stable oxygen isotopes as measured in the GISP2 deep ice core
  from Greenland. The time series is divided in three parts:  The early ice age,
  the late ice age and the Holocene. Data from \cite{braun2} and  \cite{gisp2}.} 
 
\end{figure}

Fig.~\ref{Fig1Sebas} shows the reconstructed Greenland temperature based on the
ratio of two stable oxygen isotopes
 as measured in the GISP2 deep ice core up to about $80,000$
years before present. Note that the dating accuracy typically decreases with
increasing age of the ice. The temperature-proxy obtained from these records
\cite{gisp2}  reveals at least three different dynamical regimes: the early
ice age, the late ice age and the holocene. In a simplified approach, these
regimes can be interpreted as the result of the switching between two states
(a warm one and a cold one) driven in part by noise. The interval between $12,000$
and $50,000$ years before present (late ice age) would correspond  to the ghost stochastic resonance regime, whereas the last $10,000$ years (known as holocene) stand for an age without transitions.

The ideas underlying this model are the same as those presented in
section \ref{sec:D-O}: DO events represent transitions between
two different climate states corresponding to the glacial cold (i.e., stadial)
and the glacial warm (i.e., interstadial) modes of the North Atlantic
thermohaline circulation. Note that the Holocene mode of the ocean circulation
is assumed to correspond to the glacial interstadial mode. The assumptions in the model can be summarized as follows:
\begin{enumerate}
\item The existence of two states, the glacial cold and the glacial warm (or, holocene) ones.
\item The states represent different modes of operation of the thermohaline circulation  in the North Atlantic region.
\item The  model is forced by a periodic input with frequencies close to the
  leading spectral components of the reported De Vries/Suess and Gleissberg
  cycles ($\sim 207$ and $\sim 87$ years respectively) plus a stochastic component.
\item A transition between the states takes place each time a certain threshold is crossed.
\item After each transition the threshold overshoots and afterwards approaches equilibrium following a millennial time scale relaxation process.
\item During the Holocene the periodic forcing is not able by itself to produce transitions, and the climate system remains in the warm state.
\end{enumerate}

The model is presented as a dynamical system submitted to a periodic forcing (with the frequencies mentioned before) in a double well potential and a stochastic component representing non-periodic forcing components.
It is described by the following set of differential equations:

\begin{eqnarray}
\dot{x} &=& \frac{1}{a}\left[ y.(x-x^3)+f(t) + D\sqrt a \xi(t) \right ] \\
\dot{y} &=& -\frac{y}{\tau_s}+\delta_s,
\label{eq1}
\end{eqnarray}

where $y.(x-x^3)=-\frac{dV}{dx}$ and $V(x)$ is a double well potential  with a
potential barrier following the dynamics of the $y$
variable. $f(t)=F.\left(\cos(2\pi f_1t+\phi_1)+\alpha \sin(2\pi
  f_2t+\phi_2)\right ))$ mimics the solar forcing with $f_1= 207$years$^{-1}$
and $f_2=87$years$^{-1}$. The term $D\xi(t)$ stands for a white noise process
of zero mean and amplitude $D$, and $a$ is a scaling constant. In the equation
for the threshold dynamics, $\tau_s$ and $\delta_s$ were the characteristic time decay and asymptotic threshold respectively ($s=1(0)$ for warm (cold) state). 


\begin{figure}[htpb]

\begin{center}
\includegraphics[width=6in]{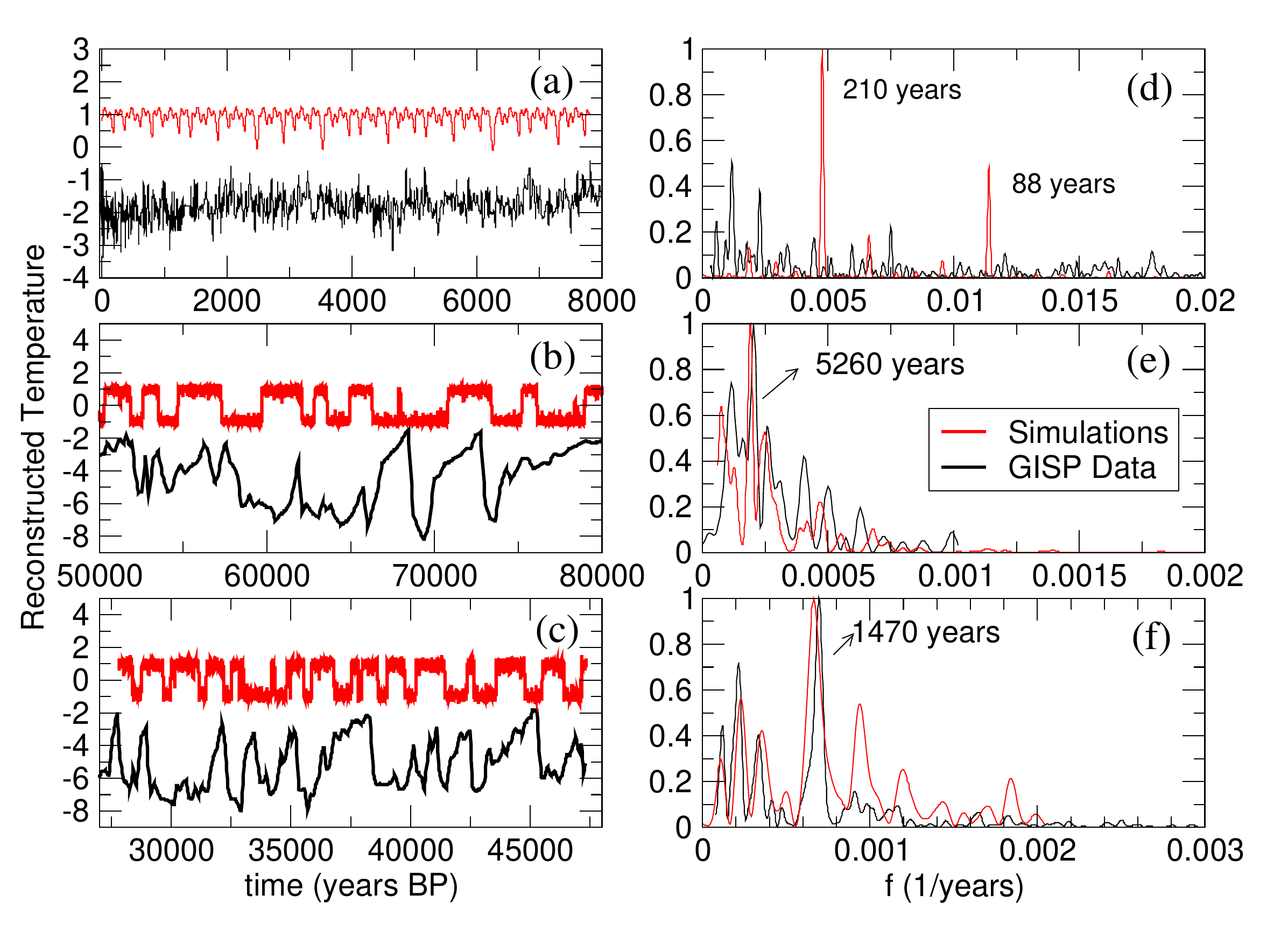}
\end{center}
\caption{\label{Fig2Sebas} Comparison between reconstructed temperature based
  on the ratio of two stable oxygen isotopes as measured in GISP2 deep ice core from
  Greenland and variable $x$ from the simulations. Left panels: Time series for (a) Holocene, (b) Early ice age and (c) Late ice age. Right panels:  Discrete Fourier Transform from these series. Replotted from \cite{sebas}.} 
 
\end{figure}

In the absence of noise there is no transitions between states, and for an optimal amount of noise the system switches between the warm and the cold state every  $1470$ years approximately,  which
corresponds to a spectral component that is absent in the periodic input. The regimes exhibited by the model for 
different noise amplitude $D$ was explored as presented in  Fig.~\ref{Fig2Sebas} which shows a comparison between the output of the model ($x$) and the GISP2 reconstructed temperature. The comparison includes the temperature records as well as their frequency spectra obtained via a Discrete Fourier Transform (DFT). The three dynamical regimes were tuned in order to obtain the best matching between the Fourier spectra. 
Even though the comparison between both series is not good enough during the
Holocene (panels (a) and (d)), a  good agreement exists during the
ice age where the main features of the dynamics are driven by
transitions between the two states. It is important to remark that during the
late Ice Age (panels (c) and (f)), the best fit between the temperature records
and the simulations corresponds to a dynamical scenario very close to the ghost stochastic resonance.

\section{Ghost Stochastic Resonance in other systems.}

\subsection{Visual Perception}\label{SecVisual}

The robustness of the  models discussed so far, as well as the absence of fitting parameters,  suggest that GSR should be found in other sensory systems.   A phenomenon that shows the analogy between the visual and auditory systems was analyzed in \cite{visual}, where the subjective rate of flickering for compound waveforms without their fundamental frequencies were measured.
The questions in these experiments were: (I) Could the observers perceive flicker at a fundamental frequency which is absent in the stimuli? and (II) Could this perception be sustained even if the phases of the higher harmonics are randomized?

\begin{figure}[htpb]

\begin{center}
\includegraphics[width=5in]{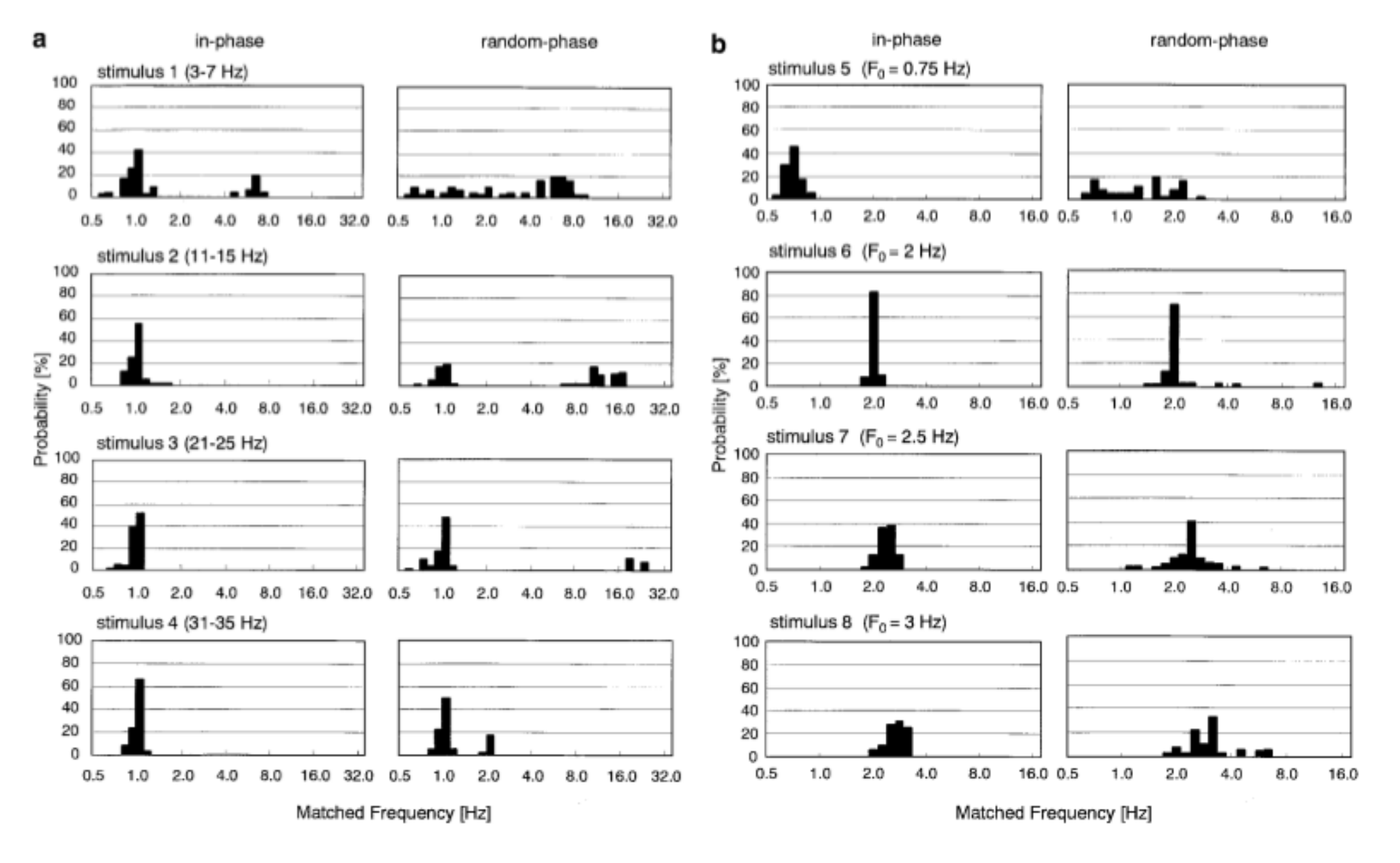}
\end{center}
\caption{\label{Fig2Visual} GSR in response to a flickering visual stimuli. Each graph represents the perceived frequency of flickering of a visual stimuli reported by four observers. The abscissa represents the frequency of the comparison stimulus matched with each test stimulus. On the ordinate, the probability of the response is represented in percentiles.  Inputs  comprised several harmonics  (indicated in the figure) with zero (denoted ``in-phase'') or random phase differences.   
With kind permission of Springer+Business Media: Vision Research 40 (2000), pp.2135-2147,  K. Fujii, S. Kita, T. Matsushima and Y. Ando. Fig.2.
} 
 
\end{figure}

To answer these questions  eight different stimuli driving a light emitting
diode were presented to a group of subjects. All the stimuli were complex
waveforms consisting of five components. The frequency of each component
corresponded to the $n^{th}$ harmonic of the common $f_0$. These components
had equal amplitude and were added to construct two kinds of waveforms for
each stimulus:``In-phase" and ``random-phase" waveforms for five different
values of $f_0$ ($f_0=0.75-3$Hz). Further details of the experimental
conditions can be found in \cite{visual}. Fig.\ref{Fig2Visual} shows the main
results of these experiments in which the observers judged the flicker
frequency in comparison with the flickering of a test stimuli composed by a single
frequency.  Note that in all cases the subjects reported a flickering rate close to the absent $f_0$, even in the case of random-phase stimuli.  Additional experiments, modifying the frequencies of each component, could replicate that the pitch shift effect described in previous section.

The results of these experiments suggest that a phenomenon analogous to the missing fundamental illusion can be found  in spatial vision as well. Visual patterns with the fundamental missing were already used in experiments of motion perception \cite{brown}, were  square-wave gratings without the fundamental were presented to a group of subjects. They perceived backwards motion when presented in quarter-cycle jumps (even though their edges and features all move forward).  Even tough these experiment were not directly related with GSR, it stresses the ubiquity of the missing fundamental phenomenon across physiological systems.

\subsection{Lasers}\label{SecLasers}

The uncovering of GSR phenomena in a variety of environments drove the search to other dynamical scenarios where similar phenomena could take place. Semiconductor lasers subject to optical feedback produce a rich dynamical behavior, including  important similarities with neuronal dynamics. 
One of their most interesting ones is the Low Frequency Fluctuation regime
(LFF) in which the output power of the lasers suffers sudden dropouts to
almost zero power at irregular time intervals when biased close to the threshold
\cite{ref4ghost}. It was shown \cite{ref5ghost,ref7ghost} that  before the onset of the LFFs a laser
 is stable under small periodic perturbations of bias current and exhibits the three ingredients of any excitable system, namely: the existence of a threshold for the perturbation amplitude above which the dropout events can occur; the form and size of the dropout events are invariant to changes in the magnitude of the perturbation; a refractory time exists: if a second perturbation is applied at a time shorter than the refractory time, the system no longer responds.

It has been shown both experimentally \cite{ref9ghost, ref10ghost} and
numerically  \cite{ref11ghost, ref12ghost} that a laser subject to optical
feedback can exhibit stochastic  and coherence
resonance when biased close to the threshold. In what follows, results
concerning experimental and numerical responses of a semiconductor laser
subject to optical feedback, biased close to the threshold and modulated by
two weak sinusoidal signals, are described  \cite{ghost}. Two-frequency forcing of dynamical systems has been already studied \cite{ref9ghost} with an emphasis on quasi-periodic dynamics. In contrast, these results show a resonance at a frequency that is absent in the input signals, i.e., GSR. 

The experimental setup consisted of an index-guided AlGaInP semiconductor laser (Roithner RLT6505G), with a nominal wavelength of 658 nm (further details can be found in \cite{ghost}).  The driving signal were composed of the superposition of the two immediate superior harmonics of $f_0=4.5$~MHz.
\begin{figure}
\begin{center}
\includegraphics[width=6in]{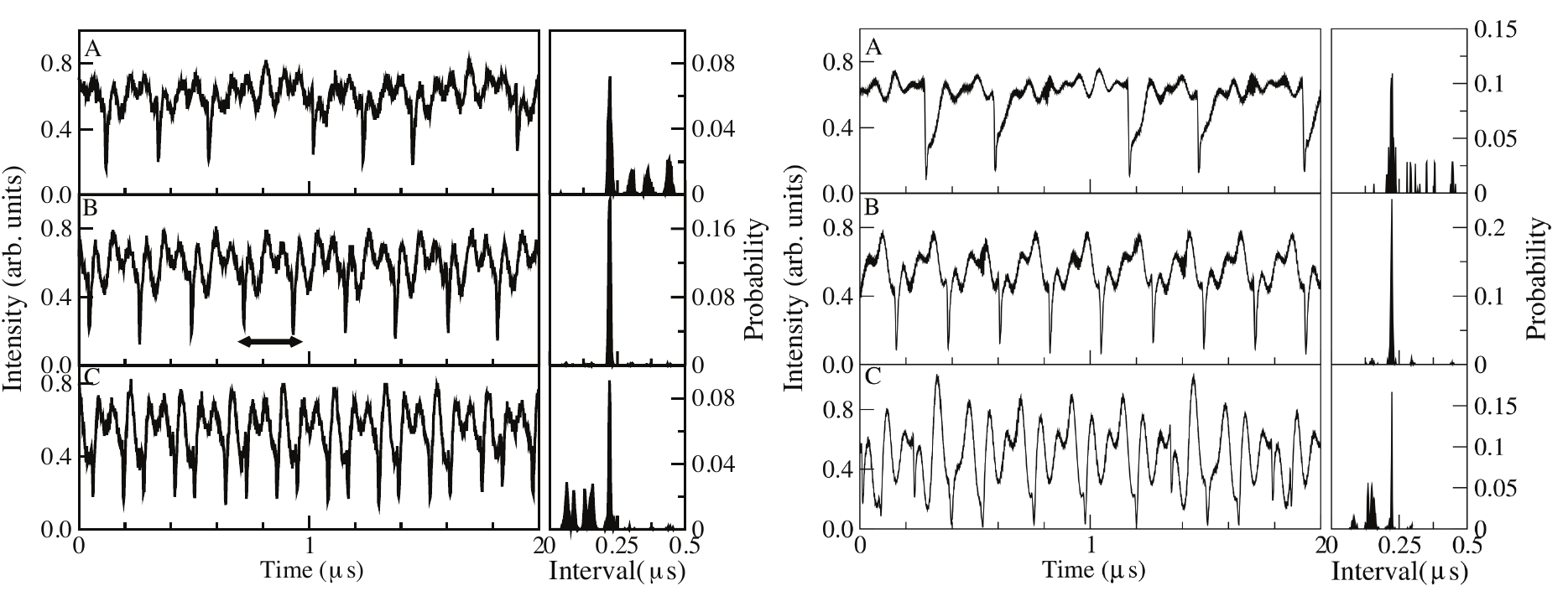}
\end{center}
\caption{ \label{Fig2LaserBuldu}  GSR in lasers. Left (right) panels: experimental (numerical) results. Time series of the optical power
in response to low (A), medium (B) and high (C) amplitudes of the injected signals. The probability distribution functions  (PDFs) of the dropouts intervals at the three amplitudes are also shown. The PDFs largest peak corresponds to $1/f_0$. In all cases the driving signal contains two frequencies. Reproduced from ref.\cite{ghost}.}
\end{figure}

Fig.~\ref{Fig2LaserBuldu} shows representative time traces and probability distribution functions (PDF) of dropout events. The left plot of the figure corresponds to experimental data for low, intermediate and high  amplitude values of the injected signals. It can be clearly seen that for an intermediate amplitude the dropouts are almost equally spaced at a time interval that corresponds grossly to $1/f_0$ (depicted by the double-headed arrow in the middle panel), a frequency that is not being injected. Thus the laser is detecting the subharmonic frequency in a nonlinear way. To better visualize this fact, the PDFs for a large number of dropouts (approximately 1500) is plotted. For the small amplitude (top-right panel in each side)  a peak at a time $1/f_0$  can be observed.  Also there were other peaks at longer times, which indicate that the system responds sometimes to $f_0$ although at some others times dropouts are skipped. For the optimum value of the amplitude (middle-right panel in each side) the PDF has a clear peak at $1/f_0$ indicating  that the system is resonating with this frequency. For the higher amplitude (bottom-right panel on each side), there are several peaks at different times corresponding to higher frequencies.
\begin{figure}
\begin{center}
\includegraphics[width=5in]{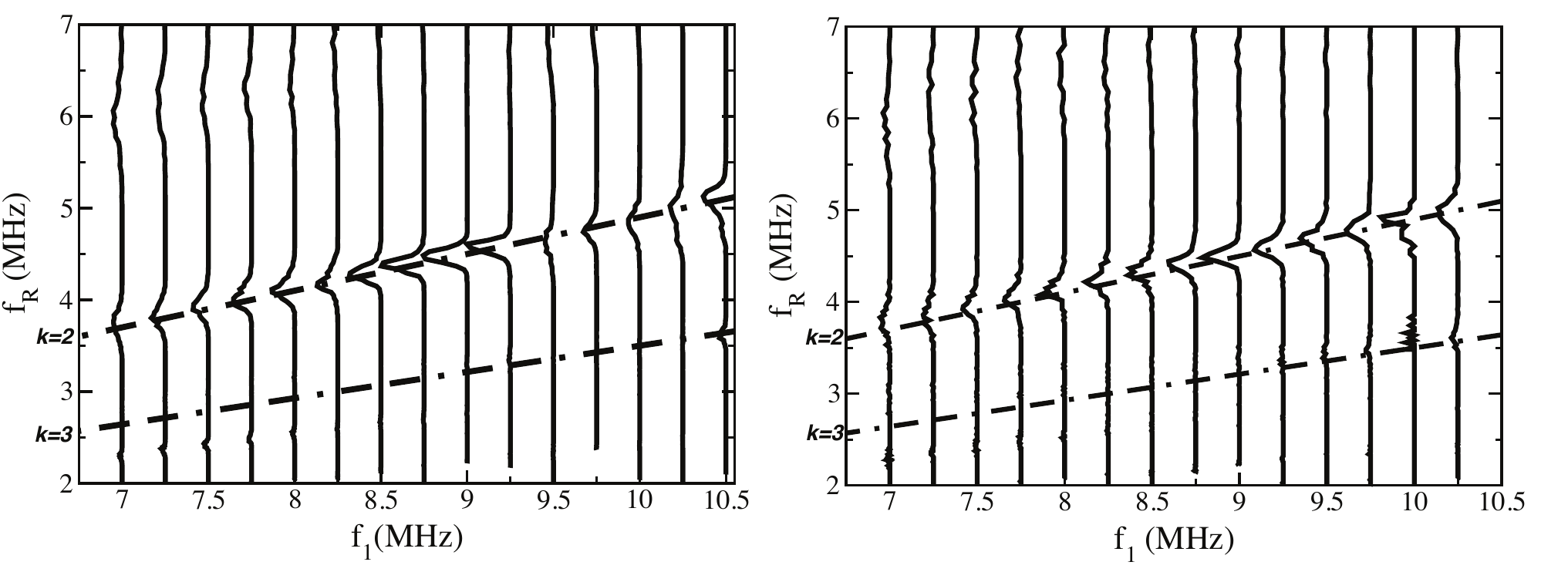}
\end{center}
\caption{ \label{Fig4LaserBuldu}  GSR  in lasers, inharmonic case. Right (left) side: experimental (numerical) results. PDFs of the intervals between dropouts are plotted as their inverse. For each pair of driving $f_1-f_2$ frequencies explored, the resulting PDF is plotted at the corresponding $f_1$ frequency. The lines are the expected resonance frequencies from the theoretical prediction given in Eq.\ref{Deltaf}. Reproduced from ref.\cite{ghost}.}%
\end{figure}

Fig.\ref{Fig4LaserBuldu} shows the results of the dropouts statistics when
both input frequencies are  shifted by the same quantity $\Delta f$. Experiments revealed how the resonant frequency of the laser followed the dynamics predicted by Eq.\ref{Deltaf}, supporting the robustness of the proposed mechanism.

Similar results were found in \cite{bghost} for two coupled lasers driven separately by a distinct external perturbation each, and show that the joint system can resonate at a third frequency different from those of the input signals. In other words, the GSR in this case was mediated by the coupling between the dynamical elements. Even though these experiments were performed in the excitable regime of the semiconductor lasers, similar results were found when studying the polarization response of a vertical-cavity surface-emitting laser, driven simultaneously by two (or more) weak periodic signals in the bistable regime \cite{sande05}, confirming  the occurrence of GSR. 

\subsection{Electronic Circuits}\label{SecCirc}

GSR was also explored  in electronic circuits whose dynamical behavior emulate neuronal dynamics. In what follows, two different configurations using  Monostable Schmitt Trigger and Chua circuits are analyzed.

\subsubsection{Monostable Schmitt Trigger.}

The behavior of a neuronal-like electronic circuit  was explored in response to a complex signal plus noise  \cite{oscar}. 
The  system considered was a non-dynamical threshold device \cite{gingli}, which compares a complex signal $S_c$ with a fixed threshold, and emits a ``spike'' ( i.e. a rectangular pulse of relatively short fixed duration) when it is crossed from below. This behavior emulates, in a very simplified way, the neuronal``firing''. The complex signal $S_c$ is formed by adding pure tones with frequencies $f_1=kf_0$, $f_2=(k+1)f_0$, ...., $f_n=(k+n-1)f_0$ plus a zero mean Gaussian distributed white noise term. The circuit  implementing the threshold device was comprised by two monostable Schmitt Triggers and up to five input frequencies combinations (i.e. $n \le 5$) were explored. The output of the circuit was digitized  and processed offline to compute intervals of time between triggering, from which an inter-spike interval (ISI) histogram  was calculated. The signal to noise ratio (SNR) was computed as before: The ratio between the number of spikes with ISI equal to (or near within $\pm5\%$) the time scale of $1/f_0$, $1/f_1$ and $1/f_2$, and the total number of ISI (i.e. at all other intervals). 


\begin{figure}[htpb]
\begin{center}
\includegraphics[width=5in]{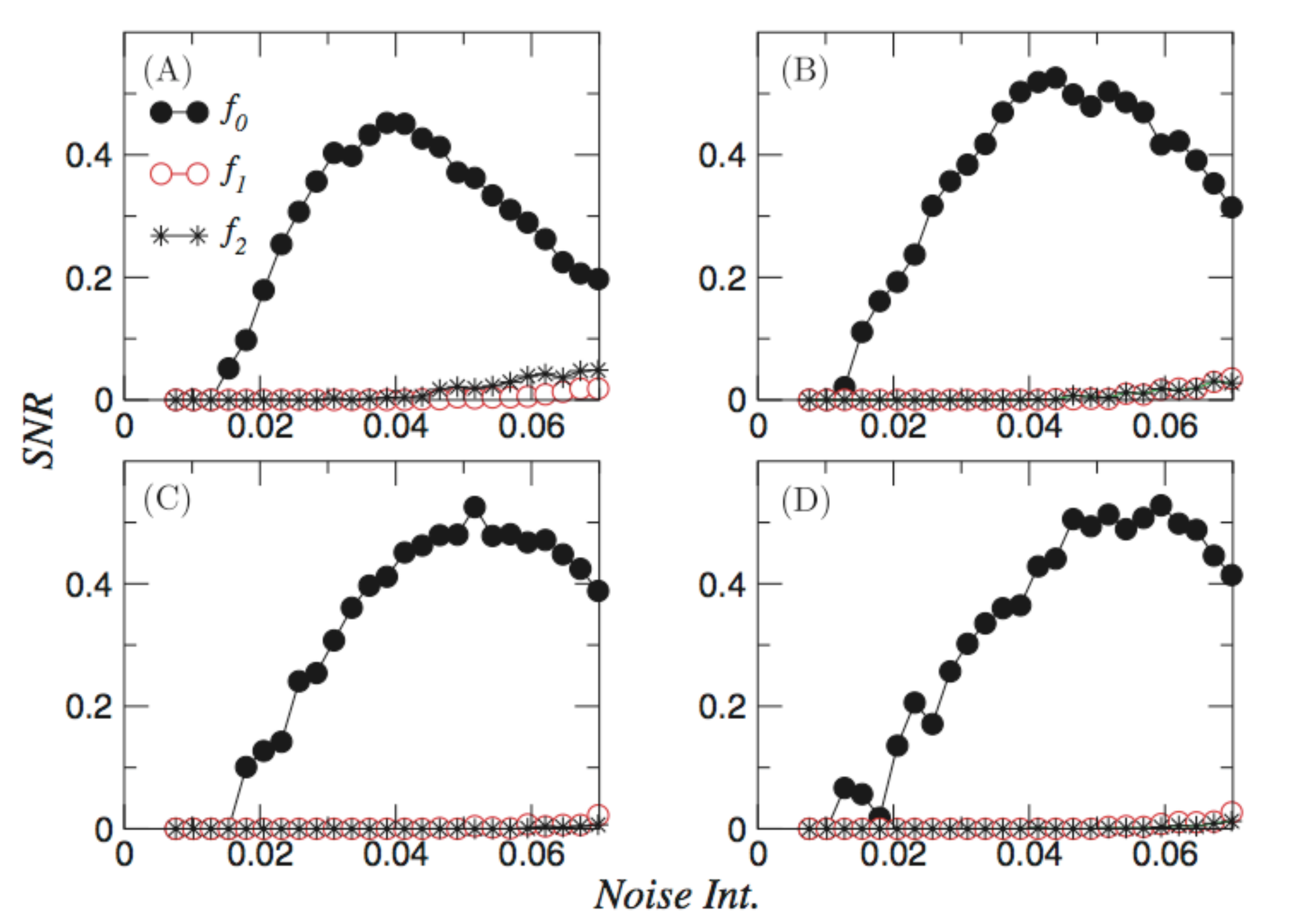}
\end{center}
\caption{\label{Fig2Calvo} GSR  in a monostable Schmitt Trigger . The figure shows   the signal-to-noise ratios versus the noise intensity for signals with two
  to five frequencies (panels A to D respectively). These are computed as the probability of observing an inter-spike interval close to the time scales (with a $5\%$ tolerance) of the frequencies $f_0$ (filled circles ), $f_1$ (empty circles ) and $f_2$ (stars ). Notice that the largest resonance is always for the ghost $f_0$, while the others are negligible.
Reprinted with permission from  O. Calvo and D.R. Chialvo, IJBC 16, pp. 731-735, (2006) . Copyright (2006) from World Scientific Publishers Co.} 
 
\end{figure}

Fig.~\ref{Fig2Calvo} shows the results from the experiments using harmonic signals composed by up to five periodic terms (i.e. $S_c$ with $n = 2, 3, 4, 5$ and $f_0 =200$~Hz).   
Even though the output was rather incoherent with any of the input frequencies
(empty circles and stars),  it is clear that it was maximally coherent
at some optimal amount  of noise, with the period close to $1/f_0$ (filled
circles). As  in the previous cases, $f_0$ was a frequency absent in the signals used to drive the system, demonstrating another instance of GSR.

\begin{figure}[htpb]
\begin{center}
\includegraphics[width=5in]{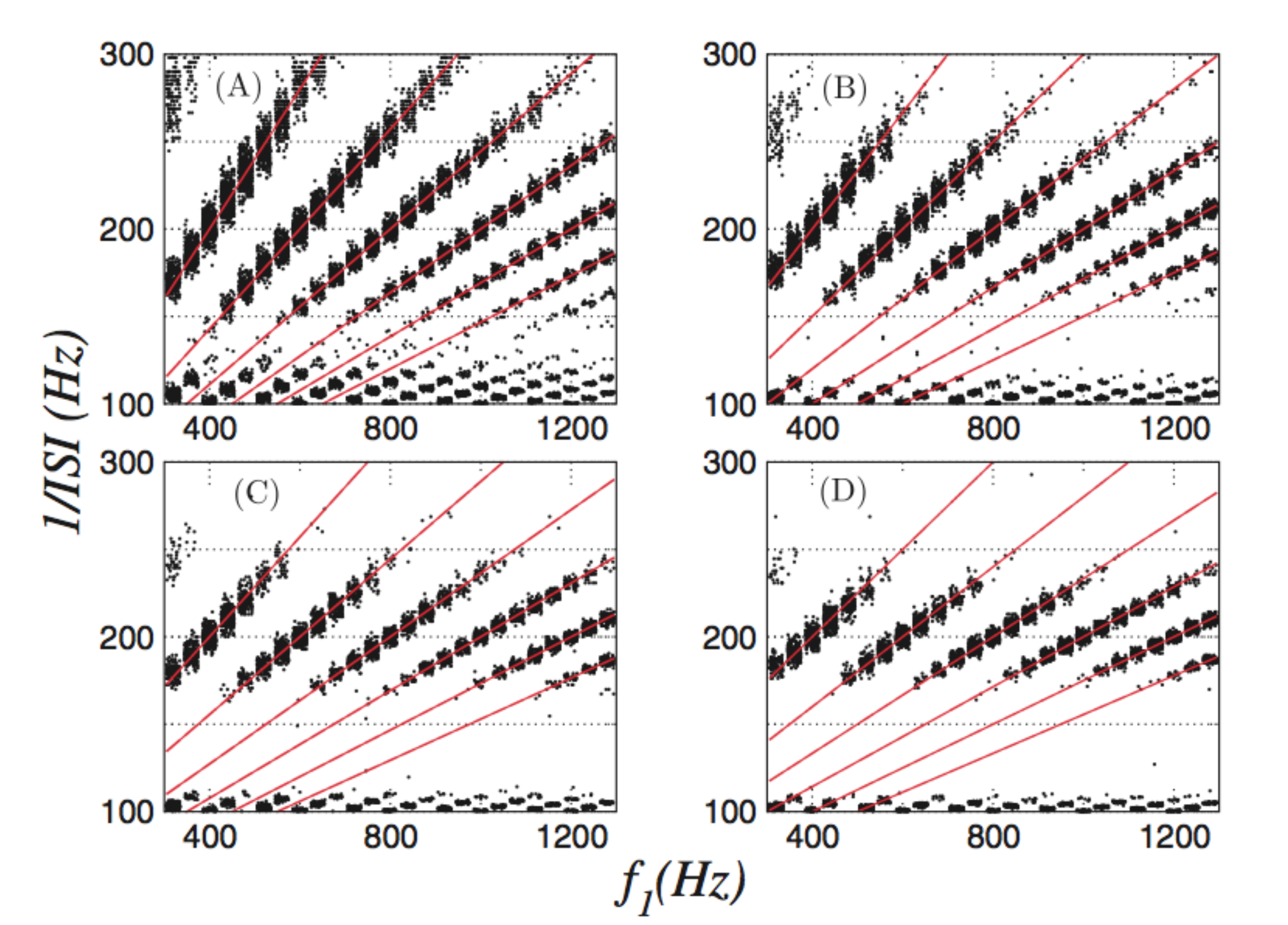}
\end{center}
\caption{\label{Fig3Calvo} Frequency shift experiments in a monostable Schmitt
  Trigger. Shown are the main resonances for signals with two to five frequencies (panels A to D respectively). In each panel, the intervals (plotted as its inverse, $f_r$) between triggered pulses  are plotted as a function of $f_1$, which was varied in steps of $40$Hz. Family of over-imposed lines are the theoretical expectation (i.e. Eq. \ref{Deltaf} with $N = 2, 3, 4 ,5$  in panels A through D respectively) for increasing $k = 2 - 7$. Reprinted with permission from  O. Calvo and D.R. Chialvo, IJBC 16, pp. 731-735, (2006) . Copyright (2006) from World Scientific Publishers Co.} 
\end{figure}

The effects of frequency shift in the harmonic inputs of $S_c$ were also explored in this circuit.  The
response of the circuit is plotted in Fig.\ref{Fig3Calvo}. As it was observed
in the previous sections, the agreement between the experimental results and
the theoretical predictions is remarkable.

\subsubsection{Pulsed coupled excitable ``Chua" circuits}

In Section \ref{SecBinau}, simulations for a binaural configuration of GSR in numerical simulations were discussed. Here,  results where the same mechanism is explored experimentally via pulsed coupled electronic neurons \cite{lopera06} are shown. To that end, two excitable electronic circuits were driven by different sinusoidal signals producing periodic spike trains at their corresponding frequencies.  Their outputs plus noise were sent to a third circuit that processed these spikes signals.  

The model is the electronic implementation of the so-called Chua circuit \cite{ref15Lopera} in the excitable regime. The two input circuits were harmonically driven at two different frequencies, $f_1$ and $f_2$, generated by a wave generator. The amplitudes of both signals were set above the threshold of the excitable circuits in order to produce periodic spiking at their outputs. These spikes were then fed to a third processing circuit via a voltage follower (which guarantees unidirectional coupling) and an electronic adder. The later also received a broadband noisy signal. The two harmonic inputs had frequencies $f_1=kf_0+\Delta f$ and $f_2=(k+1)f_0+\Delta f$ as  explored already along these notes.

\begin{figure}[!ht]
\begin{center}
\includegraphics[width=6in]{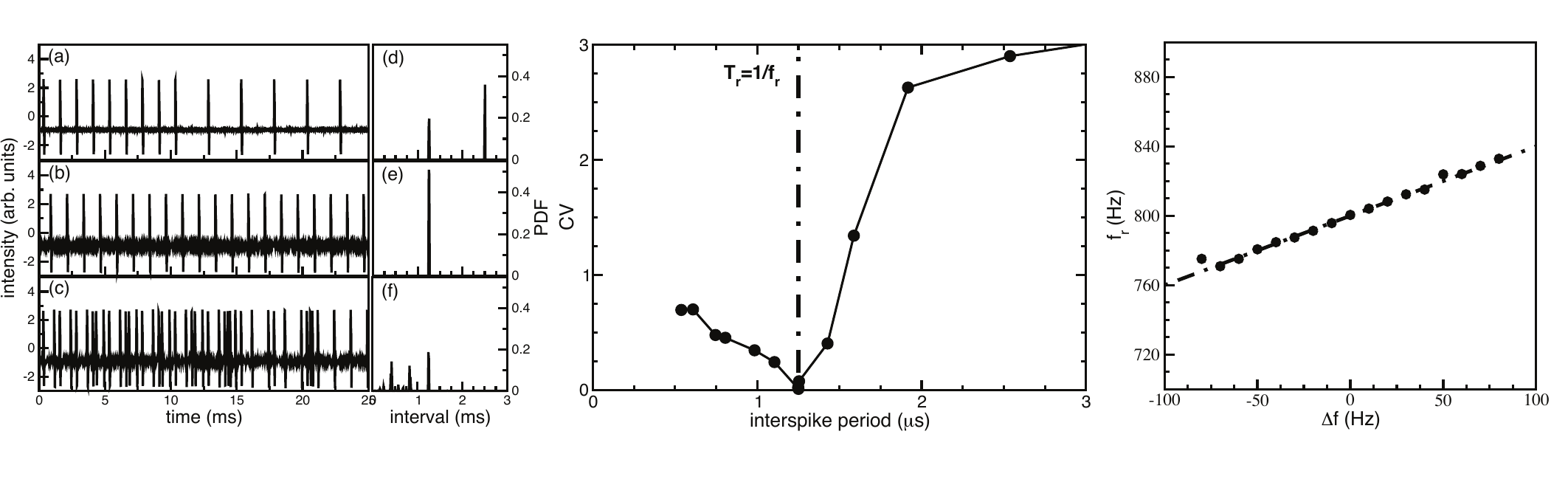}
\end{center}
\caption{GSR  in a binaural configuration of electronic neurons. Left Panels: Influence of the noise intensity on the spiking behavior of the system. The left-column plots show time series for: (a) low , (b) optimal, and (c) high values of noise . Plots (d)-(f) are the corresponding probability distribution functions of the interval between spikes. Intermediate values of noise intensity ((b), (e)) show an entrainment of the system at the ghost frequency $f_r=800$Hz ($T_r=1/f_r=1.25$ ms). Middle Panel: Coefficient of Variation (CV) of  the inter-spike interval vs its mean. The different measurements correspond to increasing values of noise. The minimum of the CV corresponds to the entrainment of the system at the ghost period ($T_r =1/f_r=1.25$ ms). Right panel: Mean spike frequency of the processing circuit for varying frequency shift  ($\Delta f$). The dashed line corresponds to the theoretical value of $f_r$ given by Eq. \ref{Deltaf} for k=2 and N=2. Reprinted with permission from A. Lopera, J. M. Buld\'u, M. C. Torrent, D. R. Chialvo, and J. Garc\'ia-Ojalvo, Phys. Rev. E 73, pp. 021101-06, (2006) . Copyright (2006) from American Physical Society.}
\label{Fig3Lopera}
\end{figure}

The frequencies of the input signals  were  $f_1=1600$Hz, $f_2=2400$Hz for the harmonic case. Left columns of Fig.~\ref{Fig3Lopera}
shows the time series of the output signal for three different values of noise
amplitudes: low, optimal and high. The corresponding histograms for the
inter-spike intervals  are displayed besides the time series. The middle panel of Fig.~\ref{Fig3Lopera} shows the coefficient of variation (CV) as a function of the mean value of inter-spike time intervals. Both panels show that there is an optimal amount of noise for which the systems responds at the missing fundamental frequency of the inputs, i.e., GSR.

As in the previous work, the influence of detuning the input frequencies was
further explored in these experiments. It means that the differences between
$f_1$ and $f_2$ was fixed in $f_0$ but they were not longer superior harmonic
of $f_0$, given that $f_1=kf_0+\Delta f$ and $f_2=(k+1)f_0+\Delta f$. The
right panel of Fig.~\ref{Fig3Lopera} shows two interesting results: one is
that the experimental data follow the theoretical prediction for $k=2$ and
$N=2$; the second is that there is not ambiguity in resonant frequencies as
was reported in the previous sections. The reason behind this difference is
related to the shape of the periodic inputs (See details in \cite{lopera06}.). 
 
\section{Beyond Ghost Stochastic Resonance}

Along these notes we have reviewed how a nonlinear system responds to a combination of pure tones with different frequencies. In all the exposed cases, the input frequencies followed a particular relation: the difference between them was always constant and equal to $f_0$, which in the harmonic case corresponds to the missing fundamental. In all cases, the analyzed systems  responded with a preferred frequency which was absent in the input.

The dynamics of  nonlinear devices, when stimulated by  more than one
frequency plus noise, was also studied from related perspectives. 
For example, the response of a discrete model system to a dichromatic input in
the regimes of stochastic and vibrational resonance was numerically analyzed
in  \cite{bicro}. The transition between the stochastic resonance regime
(where frequencies are of the same order) to the vibrational resonance one
(where one of the frequencies is much higher that the other) was studied  in bistable and threshold devices. Similar analysis
were carried out experimentally in bistable Schmitt Trigger circuits
\cite{bibi}, where the authors analyzed  the phenomena of mean switching
frequency locking and stochastic synchronization and their dependence on the input parameters.

Another perspective in the study of the dynamic of nonlinear devices driven by
more than one frequency  was carried out in \cite{rachets},  where the problem
of transport in a noisy environment was studied. In this work,  ratchet
devices were stimulated by two periodic signals with frequencies $f_1$ and
$f_2$ following rational ratios (i.e. $f_1/f_2=m/n$). The results focused  on
how the rectification of a primary signal by a ratchet could be controlled
more effectively if a secondary signal with tunable frequency and phase is applied.

\section{Open questions and future work}

Having reviewed the main theoretical aspects of  the GSR mechanism and how it contributed to explain real-world problems of active scientific research in such diverse branches as paleoclimatology and neuroscience, open questions certainly remain in at least three aspects of the phenomenon: theoretical insights, new manifestations of the same mechanism and developing of new statistical measures.

It is clear that important efforts are needed to attain analytical descriptions of the GSR mechanism, as was done earlier for the case of Stochastic Resonance. A second line of inquire might be directed to uncover new manifestations of the same mechanism. For instance, to study the characteristics of the GSR phenomenon in response to more realistic, non-sinusoidal cycles, since many real-world cyclic processes are far from sinusoidal. Another relevant question may be to investigate the response to input frequencies which are not constant but jitters randomly from one cycle to another, such that noise appears in the  frequency rather than in the amplitude domain. Furthermore, it is important to note  that most of the studies performed so far on the subject of GSR were based on comparably simple, low-dimensional models. Since it is not certain up to what degree the complexity of real-world systems (such as the climate or auditory systems) can be reduced, future work should focus on the study of this phenomenon in more detailed, higher dimensional models.
Finally, future work is required concerning the development of new measures of regularity that are particularly useful to distinguish between a GSR and other related noise induced phenomena in real data. This includes, among other approaches, rigorous null-hypothesis testing, using modern methods of non-linear time series analysis.
\section{Summary}

The output of nonlinear systems driven by noise and periodic  stimulus with more than one frequency is characterized by the emergence of a ``ghost" frequency which is absent in the input. This phenomenon, called ghost stochastic resonance, was proposed to explain a well known paradox in psychoacoustic: the missing fundamental illusion. It was later found  to provide a theoretical framework to  understand a wide variety of problems, from the perception of pitch in complex sounds or visual stimuli to  climate cycles. The robustness of this phenomenon relies in two simple ingredients which are necessary to the emergence of GSR:  the linear interference of the periodic inputs and  a nonlinear detection of the largest constructive interferences, involving a noisy threshold.
Theoretical analysis showed that  when the input frequencies are higher harmonics of some missing  fundamental $f_0$, the predominant output frequency is $f_0$. On the other hand, when the input frequencies are still spaced by $f_0$ but they are no longer higher harmonic of $f_0$, the predominant response frequencies follow a family of linear functions described by Eq.\ref{Deltaf}. The remarkable agreement found between theory, simulations and experiments is parameter independent and able to explain problems in a wide variety of systems ranging from neurons, semiconductor lasers, electronic circuits to models of glacial climate cycles.

\subsection*{Acknowledgements}
The work was supported by CONICET (grant No. PIP:0802/10)  and UBACyT (grant
No. 0476/10)(Argentina), and by the DFG (grant No. BR 3911/1)(Germany). The authors wish to dedicate this article to the memory of Prof. Frank Moss who championed, both in enthusiasm and originality, the field of stochastic resonance.

\label{lastpage}

\end{document}